\documentclass[aps,superscriptaddress,showpacs,pre,10pt,nofootinbib,twocolumn]{revtex4-1}
\usepackage{bm} 
\usepackage{graphicx} 
\usepackage{amsmath}
\usepackage{amsthm}
\usepackage{amssymb} 
\usepackage{epstopdf} 
\usepackage{amsfonts}
\usepackage{epsfig}
\usepackage{color} 
\usepackage{subfigure}
\usepackage[english]{babel}
\usepackage[bb=boondox]{mathalfa}
\usepackage{mathtools}
\usepackage{hyperref}
\hypersetup{
    colorlinks=true,       
    linkcolor=cyan,          
    citecolor=magenta,        
    filecolor=magenta,      
    urlcolor=cyan,           
    runcolor=cyan
}

\newcommand{\bra}[1]{\langle #1 |}
\newcommand{\ket}[1]{| #1 \rangle}

\newcommand {\be}{\begin{equation}}
\newcommand {\ee}{\end{equation}}

\newcommand\D{{\text{c}}}

\newcommand{\ba}{\begin{eqnarray}}
\newcommand{\ea}{\end{eqnarray}}
\newcommand{\cl}{\text{c}}
\newcommand\tr{{\mbox{Tr\,}}}
\newcommand{\ignore}[1]{}

\newcommand\ident{{\mathbb{1}}}

\usepackage[margin=1in,nofoot]{geometry}

\newcommand{\Tr}{{\mathrm{Tr}}}
\newcommand{\e}{{{e}}}
\newcommand{\rmd}{{\text d}}
\renewcommand{\Re}{\operatorname{Re}}

\newcommand{\beq}{\begin{equation}}
\newcommand{\eeq}{\end{equation}}
\newcommand{\beqnn}{\begin{equation*}}
\newcommand{\eeqnn}{\end{equation*}}
\newcommand{\bea}{\begin{eqnarray}}
\newcommand{\eea}{\end{eqnarray}}
\newcommand{\beann}{\begin{eqnarray*}}
\newcommand{\eeann}{\end{eqnarray*}}
\newcommand{\bes} {\begin{subequations}}
\newcommand{\ees} {\end{subequations}}

\begin{document}
\raggedbottom 
\title{Permutation Matrix Representation Quantum Monte Carlo}
\author{Lalit Gupta}
\affiliation{Department of Physics and Astronomy and Center for Quantum Information Science \& Technology, University of Southern California, Los Angeles, California 90089, USA}
\author{Tameem Albash}
\affiliation{Department of Electrical and Computer Engineering, Department of Physics and Astronomy, and Center for Quantum Information and Control, CQuIC, University of New Mexico, Albuquerque, New Mexico 87131, USA}
\author{Itay Hen}
\email{itayhen@isi.edu}
\affiliation{Department of Physics and Astronomy and Center for Quantum Information Science \& Technology, University of Southern California, Los Angeles, California 90089, USA}
\affiliation{Information Sciences Institute, University of Southern California, Marina del Rey, California 90292, USA}

\date{\today}

\begin{abstract}
\noindent We present a quantum Monte Carlo algorithm for the simulation of general quantum and classical many-body models within a single unifying framework. The algorithm builds on a power series expansion of the quantum partition function in its off-diagonal terms and is both parameter-free and Trotter error-free. In our approach, the quantum dimension consists of products of elements of a permutation group. As such, it allows for the study of a very wide variety of models on an equal footing. 
To demonstrate the utility of our technique, we use it to clarify the emergence of the sign problem in the simulations of non-stoquastic physical models. 
We showcase the flexibility of our algorithm and the advantages it offers over existing state-of-the-art by simulating transverse-field Ising model Hamiltonians and comparing the performance of our technique against that of the stochastic series expansion algorithm. We also study a transverse-field Ising model augmented with randomly chosen two-body transverse-field interactions.
\end{abstract}

\maketitle

\section{Introduction}
Quantum Monte Carlo (QMC) algorithms~\cite{Landau:2005:GMC:1051461,newman} are extremely useful for studying equilibrium properties of large quantum many-body systems, with applications ranging  from superconductivity and novel quantum materials~\cite{Qchem1,Qchem2,Qchem3}  through the physics of neutron stars~\cite{1742-6596-529-1-012012} and quantum chromodynamics~\cite{PhysRevLett.83.3116,PhysRevE.49.3855}.  The algorithmic development of QMC remains an active area of research, with the dual goal of extending the scope of QMC applicability and improving convergence rates of existing algorithms in order to facilitate the discovery of new phenomena~\cite{sandvik:92,sandvik:99,prokofiev:98}. 

While QMC algorithms have been adapted to the simulation of a wide variety of physical systems, different models typically require the development of distinct model-specific update rules and measurement schemes. A notable recent example is the transverse-field Ising model (TFIM), which traditionally includes only single-body $X$ terms (the transverse field), supplemented with two-body $X$ terms.  While the updates associated with single-body $X$ terms can be implemented by local (in space) updates, the inherently non-local nature of the two-body $X$ terms requires novel cluster updates~\cite{troyerXX}.  Thus, a proper treatment of the Hamiltonian with single-body \emph{and} two-body $X$ terms requires updates that are different than if the Hamiltonian included only single-body or only two-body $X$ terms.
 
In this paper, we provide a QMC scheme that has the flexibility to simulate a broad range of quantum many-body models. The technique we propose here builds on a power-series expansion of the canonical quantum partition function about the classical partition function first introduced in Refs.~\cite{ODE,ODE2}).  While strongly inspired by earlier methods that expand the partition function in powers of the inverse-temperature~\cite{handscomb1,handscomb2,Sandvik1991,sandvik:92}, our expansion is more accurately described as an expansion in off-diagonal operators of the Hamiltonian.

Our formalism enables a very general treatment of Hamiltonians, allowing us to develop a QMC scheme that is applicable to a wide variety of models, ranging from highly interacting models with multi-body terms to non-interacting ones and from strongly quantum models to purely classical ones, using the same updating formalism.  
In order to demonstrate the advantage of the new method, we study the TFIM with random two-body longitudinal interactions, comparing the performance of our technique against that of stochastic series expansion QMC~\cite{sandvik:03}. In addition, we study in detail a variant of the model with added random $XX$ interactions.  This model poses a challenge for traditional QMC approaches~\cite{doi:10.1063/1.437829,troyerXX} due to the random connectivity of the Ising couplings and $XX$ interactions, which vary between instances. 
While we focus most of our attention in what follows on finite-dimensional Hamiltonians, the technique we present here should apply with equal rigor to infinite-dimensional systems. 

The paper is organized as follows. In Sec.~\ref{sec:perMatRep}, we describe the `permutation matrix representation' (PMR) of Hamiltonians on which the partition function expansion detailed in Sec.~\ref{sec:par} is founded. We discuss the emergence of the sign problem within the formulation and its sometimes-intricate relation with the concept of non-stoquasticity in Sec.~\ref{sec:ns} and in Sec.~\ref{sec:QMCalgo} we present the QMC algorithm we have devised based on the expansion. 
Results of TFIM simulations are presented in Sec.~\ref{sec:Results}. We conclude in Sec.~\ref{sec:conclusions} with additional discussions and some caveats.


\section{The permutation matrix representation\label{sec:perMatRep}}

We consider many-body systems whose Hamiltonians we cast as the sum
\beq \label{eq:basic}
H=\sum_{j=0}^M \tilde{P}_{j} =\sum_{j=0}^M D_j P_j =H_c + \sum_{j=1}^M D_j P_j \,,
\eeq
where $\{ \tilde{P}_j\}$ is a set of $M+1$ distinct generalized permutation matrices~\cite{gpm}, i.e., matrices with precisely one nonzero element in each row and each column (this condition can be relaxed to allow for rows and columns with only zero elements). Each operator $\tilde{P}_j$ can be written, without loss of generality, as $\tilde{P}_j=D_j P_j$ where $D_j$ is a diagonal matrix\footnote{The diagonal matrix $D_j$ will be invertible, i.e., will not contain zero elements along the diagonal, if $\tilde{P}_j$ is a bonafide generalized permutation matrix.} and $P_j$ is a  permutation matrix with no fixed points (equivalently, no nonzero diagonal elements) except for the identity matrix $P_0=\mathbb{1}$. We will refer to the basis in which the operators $\{D_j\}$ are diagonal as the computational basis and denote its states by $\{ |z\rangle \}$. We will call the diagonal matrix $D_0$ the `classical Hamiltonian' and will sometimes denote it by $H_{\cl}$.
The permutation matrices appearing in $H$ will be treated as a subset of a permutation group, wherein $P_0$ is the identity element. 

The $\{D_j P_j \}$  off-diagonal operators (in the computational basis) give the system its  `quantum dimension'.  Each term $D_j P_j $ obeys 
$D_j P_j | z \rangle = d_j(z') | z' \rangle$ where $d_j(z')$ is a possibly complex-valued coefficient and $|z'\rangle \neq |z\rangle$ is a basis state.
While the above formulation may appear restrictive, we show in Appendix~\ref{app:PermutationM} that any finite-dimensional matrix can be written in the form of Eq.~\eqref{eq:basic}. 

We also note that $H=\sum_j D_j P_j$ is hermitian if and only if for every index $j$ there is an associated index $j'$ such that $P_j = P_{j'}^{-1}$ and $D_j = D_{j'}^*$ where the indices $j$ and $j'$ can be the same (see Appendix \ref{app:Hermicity_proof}). This in turn implies that any Hamiltonian $H$ can be written as
\beq
H=\sum_{j} R_j \left(\e^{i \Phi_j} P_j + \e^{-i \Phi_j} P^{-1}_j\right)\,,
\eeq
where $R_j, \Phi_j$ are real-valued diagonal matrices. In the case where a permutation matrix $P_j$ is its own inverse, the corresponding $\Phi_j$ will necessarily be the zero matrix.

To further elucidate PMR, we now provide several examples. 
\subsection{Example I: A single spin-$1/2$ particle}\label{sec:single_spin_ex}

The Hamiltonian of a single spin-$1/2$ particle can most generally be written as
\beq
H=\alpha_0 \mathbb{1} +\alpha_1 X + \alpha_2 Y + \alpha_3 Z \ ,
\eeq
where $X, Y$ and $Z$ are the matrix representations of the usual Pauli operators in the  basis that diagonalizes the Pauli-$Z$ operator.  In PMR, the Hamiltonian becomes
\beq\label{eq:single_spin}
H=D_0 P_0+ D_1 P_1
\eeq
with $P_0 =  \mathbb{1}$, $P_1 = X$, $D_0=H_c=\alpha_0 \mathbb{1}+ \alpha_3 Z$ and $D_1=\alpha_1 \mathbb{1}-i \alpha_2 Z$.

\subsection{Example II: Two-local spin-$1/2$ models}
A general two-local $n$-particle spin-$1/2$ Hamiltonian has similarly  the following form
\beq
H = \sum_{i<j} \sum_{\substack{K_i \in \left\{\ident_i, X_i,Y_i,Z_i \right\} \\ K_j \in \left\{\ident_j, X_j,Y_j,Z_j \right\}}} \alpha_{ij,K_i,K_j} K_i K_j \,.
\eeq
Here, the basis states are tensor products of the single spin states. 
We can cast the Hamiltonian in the form of Eq.~\eqref{eq:basic} by grouping together elements that change a given basis state $|z\rangle$ to the same basis state $|z'\rangle$. For example, the terms $ X_i$, $Y_i$,  $X_i Z_j$, $ Y_i Z_j$ are grouped together as the action of the combined term \hbox{$V_i=\alpha_{ij10} X_i+ \alpha_{ij20} Y_i + \sum_j (\alpha_{ij13} X_i Z_j+ \alpha_{ij23} Y_i Z_j)$} can be written as $D_i X_i$, where $D_i$ is  a (generally complex-valued) diagonal matrix. This approach can be straightforwardly generalized to three- and higher-local Hamiltonians. 
The permutation matrices $P_i$ for general spin-1/2 Hamiltonians are by extension $\{ \mathbb{1}, X_i, \ldots, X_i X_j,\ldots,X_i X_j X_k,\ldots\}$.

\subsection{Example III: Spin-one particles (qutrits)\label{sec:qutrit}}

The permutation matrix representation generalizes straightforwardly to higher dimensional systems. The Hamiltonian for a single qutrit can be written as $H=D_0 P_0+D_1 P_1+D_2 P_2$
%
where
\[
P_0=\mathbb{1}= 
\begin{bmatrix}
    1  &  0 & 0      \\
    0 &  1 &  0	 	 \\
     0 &  0 &  1  
\end{bmatrix}\;,
P_1 = 
\begin{bmatrix}
  0  &  0 & 1      \\
    1 &  0 &  0	 	 \\
     0 &  1 &  0     
\end{bmatrix}\;,
P_2 = 
\begin{bmatrix}
  0  &  1 & 0      \\
    0 &  0 &  1	 	 \\
     1 &  0 &  0     
\end{bmatrix} \, , 
\]
and $D_2 = D_1^\ast$, a condition imposed by the hermiticity of the Hamiltonian.

\subsection{Example IV: The Bose-Hubbard model}

Another model that can just as easily be represented in permutation matrix form is the Bose-Hubbard model.
This discretely infinite dimensional model captures the physics of interacting spinless bosons on a lattice~\cite{lewenstein2012ultracold} and is commonly used to describe superfluid-insulator transitions~\cite{PhysRevB.40.546_superfluid}, bosonic atoms in an optical lattice~\cite{JAKSCH200552_optical_lattice} and certain magnetic insulators~\cite{2008NatPh...4..198G_magnetic_insulator}. 

The Bose-Hubbard Hamiltonian is given by
\beq
H = -t \sum_{ \left\langle i, j \right\rangle } \hat{b}^{\dagger}_i \hat{b}_j + \frac{U}{2} \sum_{i} \hat{n}_i \left( \hat{n}_i - 1 \right) - \mu \sum_i \hat{n}_i \,.
\eeq
Here, $\left\langle i, j \right\rangle$ denotes summation over all neighboring lattice sites $i$ and $j$, while $\hat{b}^{\dagger}_i$ and $\hat{b}^{}_i$ are regular bosonic creation and annihilation operators such that $\hat{n}_i = \hat{b}^{\dagger}_i \hat{b}_i$ gives the number of particles at the $i$-th site. The model is parametrized by the hopping amplitude $t$ and the on-site interaction $U$.

In the bosonic number basis where states are described by the number of bosons in each site $|n_1\rangle \ldots |n_{L}\rangle$ (here $L$ is the number of lattice sites) we identify the diagonal part to be
\hbox{$D_0=\frac{U}{2} \sum_{i} \hat{n}_i \left( \hat{n}_i - 1 \right) - \mu \sum_i \hat{n}_i$} and the off-diagonal (infinite dimensional) permutation operators as $P_{\left\langle i, j \right\rangle} = \hat{b}^{\dagger}_i \hat{b}_j$. The diagonal operators associated with $P_{\left\langle i, j \right\rangle}$ are $D_{\left\langle i, j \right\rangle}$ whose entries are $-t$ for states whose $n_j$ is positive (the $j$-th boson can be annihilated) and zero otherwise. 

\section{\label{sec:par}Off-diagonal partition function expansion}
We are now in a position to discuss the off-diagonal series expansion of the partition function $Z=\tr\left[ \e^{-\beta H} \right]$ as it applies to Hamiltonians cast in the form given in Eq.~(\ref{eq:basic}). 

We begin by replacing the trace operation $\Tr[\cdot ]$ with the explicit sum $\sum_z \langle z | \cdot | z \rangle$ and then expanding the exponent in the partition function in a Taylor series in $\beta$:
\bea
Z &=&\sum_{z} \sum_{n=0}^{\infty}\frac{\beta^n}{n!} \langle z | (-H)^n | z \rangle \\\nonumber 
&=& \sum_{z} \sum_{n=0}^{\infty}\frac{\beta^n}{n!} \langle z | \left(-H_\D -\sum_{j=1} D_j P_j\right)^n | z \rangle \\\nonumber
&=& \sum_{z} \sum_{n=0}^{\infty}  \sum_{\{ S_{{\bf i}_n}\}} \frac{\beta^n}{n!} \langle z | S_{{\bf i}_n} | z \rangle \, .
\eea
In the last step we have expressed $(-H)^n$ in terms of all sequences of length $n$ composed of products of basic operators $H_\D$ and $D_j P_j$, which we have denoted by the set $\{S_{{\bf i}_n}\}$. Here ${\bf i}_n=(i_1,i_2,\ldots,i_n)$ is a set of indices, each of which runs from $0$ to $M$, that denotes which of the $M+1$ operators in $H$ appear in $S_{{\bf i}_n}$. 

We proceed by stripping away all the diagonal Hamiltonian terms from the sequence $\langle z | {S}_{{\bf i}_n} | z \rangle$. We do so by evaluating the action of these terms on the relevant basis states, leaving only the off-diagonal operators unevaluated inside the sequence (see Refs.~\cite{ODE,ODE2} for a more detailed derivation). 

The partition function may then be written as
\bea\label{eq:snsq2}
Z  &=&
\sum_{z} \sum_{q=0}^{\infty} \sum_{\{ {S}_{q}\}}  \left(  \prod_{j=1}^q d^{(i_j)}_{z_j} \right) \langle z | S_{{\bf{i}}_q} | z \rangle  \left( \sum_{n=q}^{\infty} \frac{\beta^n(-1)^{n}}{n!} \right. \nonumber \\ 
&\times&\left.  \sum_{\sum k_i=n-q} (E_{z_0})^{k_0} \cdot \ldots \cdot (E_{z_{q}})^{k_{q}} \right)\,,
\eea
where $E_{z_i}=\langle z_i |H_\D | z_i \rangle$ and ${\{S_{{\bf{i}}_q}\}}$ denotes the set of all products of length $q$ of `bare' \emph{off-diagonal} operators $P_j$. 
Also 
\beq \label{eq:hsAppC}
d^{(i_j)}_{z_j} = \langle z_j | D_{i_j}|z_j\rangle \, ,
\eeq
which can be considered as the `hopping strength' of $P_{i_j}$ with respect to $|z_j\rangle$. 
Note that while the partition function is positive and real-valued, the $d^{(i_j)}_{z_j}$ elements do not necessarily have to be so. 

\begin{figure}[bp]
\includegraphics[width=.48\textwidth]{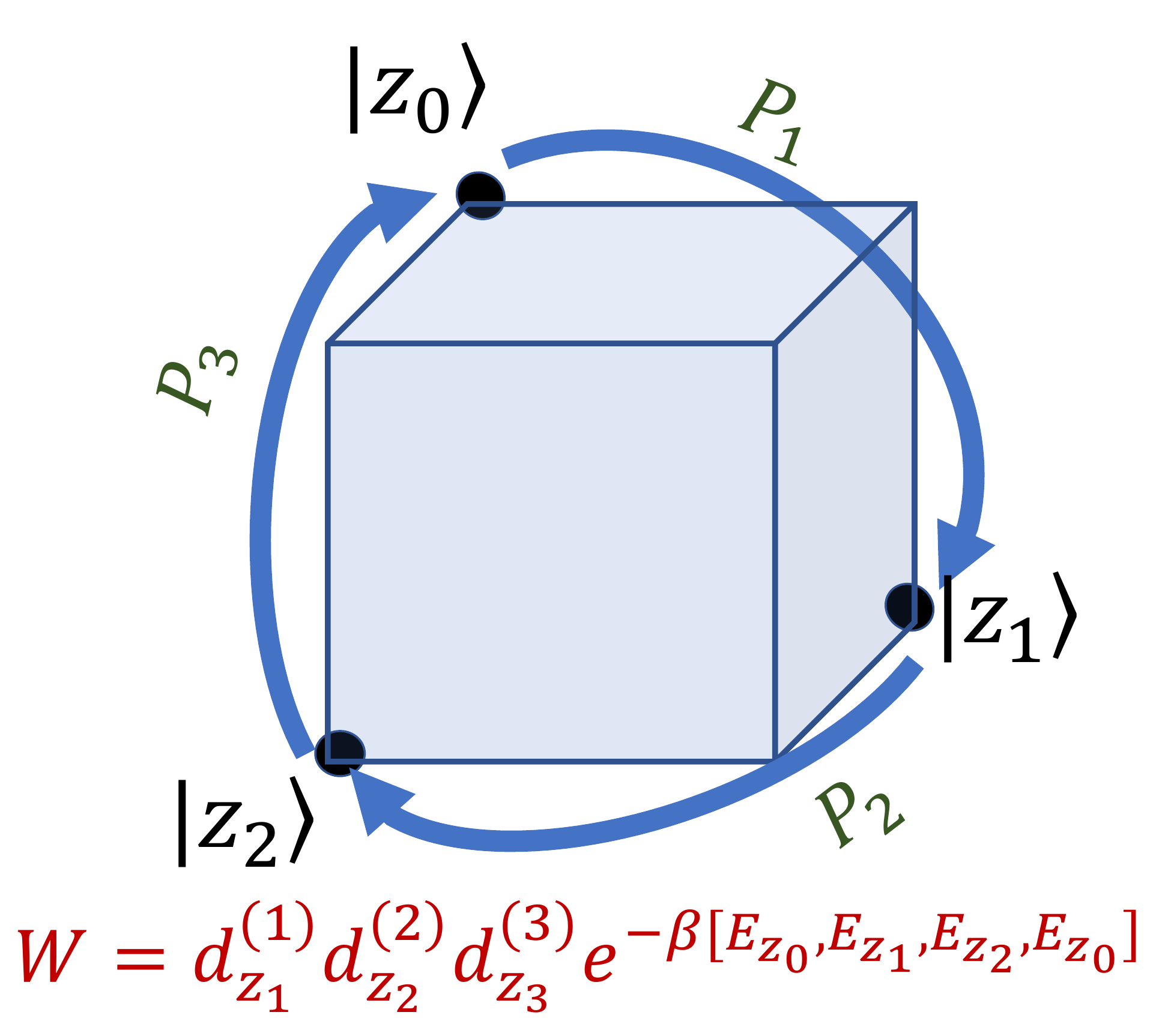}
\caption{\label{fig:hyper} Diagrammatic representation of a generalized Boltzmann weight, or a GBW, calculated from the classical energies $E_{z_j}$ of the classical states $|z_j\rangle$, which form a closed walk on the hypercube of basis states. The walk is determined by the action of the permutation operators of the configuration, represented by $S_{{\bf{i}}_q}=P_3 P_2 P_1$, on the initial basis state $|z_0\rangle$. The walk closes if and only if the sequence of permutation operators evaluates to the identity operation. 
}
\end{figure}

The term in parentheses in Eq.~\eqref{eq:snsq2} sums over the diagonal contribution of all $\langle z | S_{{\bf i}_n} | z \rangle$ terms that correspond to the same $\langle z | S_{{\bf{i}}_q} | z \rangle$ term. The various $\{|z_i\rangle\}$ states are the states obtained from the action of the ordered $P_j$ operators in the product $S_{{\bf{i}}_q}$ on $|z_0\rangle$, then on $|z_1\rangle$, and so forth. For example, for $S_{{\bf{i}}_q}=P_{i_q} \ldots P_{i_2}P_{i_1}$, we obtain $|z_0\rangle=|z\rangle, P_{i_1}|z_0\rangle=|z_1\rangle, P_{i_2}|z_1\rangle=|z_2\rangle$, etc. The proper indexing of the states $|z_j\rangle$ along the path is \hbox{$|z_{(i_1,i_2,\ldots,i_j)}\rangle$} to indicate that the state at the $j$-th step depends on all $P_{i_1}\ldots P_{i_j}$. We will use the shorthand $|z_j\rangle$. The sequence of basis states $\{|z_i\rangle \}$ may be viewed as a `walk' on the hypercube of basis states~\cite{ODE,ODE2,signProbODE} (see Fig.~\ref{fig:hyper}).

After a change of variables, $n \to n+q$, we arrive at:
\bea
Z  &=& \sum_{z} \sum_{q=0}^{\infty} \sum_{\{ {S}_{q}\}} \langle z | S_{{\bf{i}}_q} | z \rangle \left( \left( -\beta \right)^q \left(  \prod_{j=1}^q d^{(i_j)}_{z_j} \right) \right. \nonumber  \\ 
&\times& \left. \sum_{n=0}^{\infty} 
 \frac{(-\beta)^n}{(n+q)!}  \sum_{\sum k_i=n}  (E_{z_0})^{k_0} \cdots (E_{z_{q}})^{k_{q}} \right)\,.
\eea
%
Noting that the various $\{E_{z_i} \}$ are the classical energies of the states $|z_i\rangle$ states created by the operator product $S_{{\bf{i}}_q}$, the partition function is now given by:
\bea \label{eqt:infinitesum}
Z  &=& \sum_z \sum_{q=0}^{\infty} \left(  \prod_{j=1}^q d^{(i_j)}_{z_j} \right) \sum_{ \{ {S}_{q}\}}  \langle z | S_{{\bf{i}}_q} | z \rangle 
\\\nonumber&\times&
 \left( 
\sum_{\{ k_i\}=(0,\ldots,0)}^{(\infty,\ldots,\infty)} \frac{(-\beta)^q}{(q+\sum k_i)!} \prod _{j=0}^{q} (-\beta E_{z_j})^{k_j} 
\right)
 \,.
 \eea
A feature of the above infinite sum is that the term in parentheses can be further simplified to give the \emph{exponent of divided differences} of the $E_{z_i}$'s (a short description of divided differences and an accompanying proof of the above assertion can be found in Ref.~\cite{ODE}); it can therefore be succinctly rewritten as: 
\bea
\sum_{\{ k_i\}} \frac{(-\beta)^q}{(q+\sum k_i)!} \prod _{j=0}^{q} (-\beta E_{z_j})^{k_j} 
=e^{-\beta[E_{z_0},\ldots,E_{z_q}]} \nonumber\\\,, 
\eea
where $[E_{z_0},\ldots,E_{z_q}]$ is a \emph{multiset} of energies and where a function $F[\cdot]$ of a multiset of input values is defined by
\beq 
F[E_{z_0},\ldots,E_{z_q}] \equiv \sum_{j=0}^{q} \frac{F(E_{z_j})}{\prod_{k \neq j}(E_{z_j}-E_{z_k})}
\eeq
and is called the \textit{divided differences}~\cite{dd:67,deboor:05} of the function $F[\cdot]$ with respect to the list of real-valued input variables $[E_{z_0},\ldots,E_{z_q}]$. In our case, $F[\cdot]$ is the function
\beq
F[E_{z_0},\ldots,E_{z_q}] = e^{-\beta[E_{z_0},\ldots,E_{z_q}]} \,.
\eeq
Therefore, the infinite sum over energies in Eq.~\eqref{eqt:infinitesum} may be simplified to
\beq
Z  = \sum_{z} \sum_{q=0}^{\infty} \sum_{\{ {S}_{q}\}} \langle z | S_{{\bf{i}}_q} | z \rangle D_{(z,S_{{\bf{i}}_q})} e^{-\beta[E_{z_0},\ldots,E_{z_q}]} \,,
\label{eq:SSE3}
\eeq
where we have denoted 
\beq
D_{(z,S_{{\bf{i}}_q})}=\prod_{j=1}^q d^{(i_j)}_{z_j}\,.
\eeq

We stress that the partition function expanded as such is not an expansion in $\beta$.  Similar to the techniques introduced by Handcomb in the 1960s~\cite{handscomb1,handscomb2} and further developed in the stoquastic series expansion (SSE) scheme pioneered by Sandvik~\cite{Sandvik1991,sandvik:92}, the off-diagonal series expansion \emph{begins} with a Taylor series expansion of the exponential function in the inverse temperature $\beta$, but the regrouping of terms into the exponent of divided-differences means that it is no longer a high-temperature expansion.
SSE writes the trace of products of the Hamiltonian as a sum of products of matrix elements written in a suitably chosen basis which are then sampled (thereby overcoming the limitation of Handscomb's scheme which required the evaluation of traces of products of the Hamiltonian). This is made possible by breaking up the Hamiltonian into a sum of local bonds. In PMR, this is not the case --- all terms may remain non-local but are grouped according to their action on basis states. 

Specially, the diagonal portion of the Hamiltonian remains `intact' which then allows us to regroup a large portion of all terms. 
A single divided-difference term thus corresponds to a sum of an infinite number of SSE terms, meaning that a single PMR configuration represents very many standard SSE configurations and a single PMR weight sums up very many standard SSE weights.  This can be immediately seen in the derivation above, particularly Eq.~\eqref{eq:snsq2}, which relates the standard SSE weight, which involves sequences of diagonal as well as off-diagonal bonds, to the weights of the current approach that only involve off-diagonal bonds. 

It is worth noting that the cost of the massive grouping of the off-diagonal series expansion is manifested in the computational cost associated with calculating PMR terms (or ratios thereof). As we discuss in Sec.~\ref{sec:QMCalgo} in more detail, these can be calculated (or more precisely, updated) using $O(q)$ basic arithmetic operations (where $q$ is the number of operators in a sequence or equivalently the size of the imaginary-time dimension), as opposed to SSE's $O(1)$. We refer the reader to Ref.~\cite{ODE} for a more detailed comparison between the off-diagonal expansion and SSE. In Sec.~\ref{sec:bench} we provide a runtime comparison of PMR vs SSE for simulations of TFIM instances. 

Aside from SSE, it is also interesting to observe that the exponent of divided differences also has close relations to continuous-time QMC  (e.g., Ref.~\cite{prokofiev:98}), via the 
Hermite-Genocchi formula~\cite{deboor:05}:
\bea
e^{-\beta[E_0,\ldots, E_q]} = \int_{\Omega} \rmd t_0 \cdots \rmd t_{q} 
e^{-\beta\left( E_0 t_0 +E_1 t_1 + \ldots + E_q t_q\right) }  \,, \nonumber\\
\eea
where $t_i \geq 0$ and the area of integration $\Omega$ is bounded by $t_0 + t_1 + \ldots + t_{q}$ from above.

Having derived the expansion Eq.~(\ref{eq:SSE3}) for any Hamiltonian cast in the form Eq.~(\ref{eq:basic}), we are now in a position to interpret the partition function expansion as a sum of weights, i.e., $Z = \sum_{\{{\cal C}\}} W_{{\cal C}}$, where the set of configurations $\{{\cal C}\}$ is all the distinct pairs $\{ |z\rangle, S_{{\bf{i}}_q} \}$.  Because of the form of $W_{{\cal C}}$, 
\beq \label{eq:gbw}
W_{{\cal C}}=   D_{(z,S_{{\bf{i}}_q})}  \e^{-\beta [E_{z_0},\ldots,E_{z_q}]}\,,
\eeq
we refer to it as  a `generalized Boltzmann weight' (or, a GBW).  It can be shown~\cite{ODE} that $(-1)^q \e^{-\beta [E_{z_0},\ldots,E_{z_q}]}$ is strictly positive. Another feature of divided differences is that they are invariant under rearrangement of the input values.

We note that as written, the weights $W_{\cal C}$ are complex-valued, despite the partition function being real (and positive). Since for every configuration $\mathcal{C}=\{ |z\rangle, S_{{\bf{i}}_q}\}$ there is a conjugate configuration $\bar{\mathcal{C}}=\{ |z\rangle, S^{\dagger}_{{\bf i}_q}\}$\footnote{For $S_{{\bf{i}}_q} =P_{i_q} \ldots  P_{i_2} P_{i_1}$, the conjugate sequence is simply $S^\dagger_{{\bf i}_q} = P_{i_1}^{-1} P_{i_2}^{-1} \ldots P_{i_q}^{-1}$.} that produces the conjugate weight $W_{\bar{\mathcal{C}}}=\bar{W}_{\mathcal{C}}$, the imaginary contributions cancel out.  Expressed differently, the imaginary portions of complex-valued weights do not contribute to the partition function and may be disregarded altogether. We may therefore redefine \hbox{$D_{(z,S_{{\bf{i}}_q})}=\Re\left[\prod_{j=1}^q d^{(i_j)}_{z_j}\right]$}, obtaining strictly real-valued weights. 

Before we move on, we note that $\langle z| S_{{\bf{i}}_q} |z\rangle$ evaluates either to 1 or to zero.
Moreover, since the permutation matrices with the exception of $P_0$ have no fixed points, the condition $\langle z| S_{{\bf{i}}_q} |z\rangle=1$ implies $S_{{\bf{i}}_q}=\mathbb{1}$, i.e., $S_{{\bf{i}}_q}$ must evaluate to the identity element $P_0$ (note that the identity element does not appear in the sequences $S_{{\bf{i}}_q}$).  The expansion can thus be more succinctly rewritten as
 \beq\label{eq:z}
Z  =\sum_{z}\sum_{S_{{\bf{i}}_q}=\mathbb{1}}  D_{(z,S_{{\bf{i}}_q})}   \e^{-\beta [E_{z_0},\ldots,E_{z_q}]} \,.
\eeq

\section{\label{sec:ns}Non-stoquasticity and emergence of the sign problem}

An attractive property of the formalism introduced above is that it allows us to identify the emergence of the sign problem in QMC via inspection of the weights $W_{\cal C}$, thereby making more apparent the connection between the notion of non-stoquasticity --- the existence of positive or complex-valued off-diagonal Hamiltonian matrix entries --- which has garnered increasing attention with the advent of quantum computers in recent years~\cite{Bravyi:QIC08,Bravyi:2014bf,marvianLidarHen} and the onset of the sign problem. 

To interpret the real-valued weight terms $W_{\mathcal{C}}$ as actual weights (equivalently, un-normalized probabilities), they must be nonnegative. The occurrence of negative weights marks the onset of the infamous sign problem. A weight is positive iff 
$$(-1)^q D_{(z,S_{{\bf{i}}_q})}=\Re\left[\prod_{j=1}^q (-d^{(i_j)}_{z_j})\right]$$ 
is positive, that is, a QMC algorithm will encounter a sign problem, equivalently a negative weight, during a simulation if and only if there exists a closed walk on the hypercube of basis states along which $\Re\left[\prod_{j=1}^q (-d^{(i_j)}_{z_j})\right]<0$. It is thus clear that it is not mere non-stoquasticity (equivalently, the sign of off-diagonal entries) that creates the sign problem, but rather the sign of closed walks on the hypercube of basis states that determines its occurrence. 

A special class of models where the sign problem does not emerge, i.e., where $\Re\left[\prod_{j=1}^q (-d^{(i_j)}_{z_j})\right]\geq 0$ for all configurations, is that of `stoquastic' Hamiltonians~\cite{Bravyi:QIC08,Bravyi:2014bf} for which all $d^{(i_j)}_{z_j}$ are negative, which is equivalent to having only nonpositive off-diagonal elements in the matrix representation of the Hamiltonian. In this case, all products trivially yield positive-valued walks. 

The existence of positive off-diagonal terms does not however immediately imply a sign problem for QMC (the reader is referred to Ref.~\cite{elucidating} for a more detailed discussion).
Another example of a sign-problem-free family of models is one where all $d^{(i_j)}_{z_j}$ elements are positive but closed walks are all of even length. One such model is the transverse-field Ising Hamiltonian 
\beq\label{eq:Hising}
H =\sum_{i,j} J_{ij} Z_i Z_j +  \sum_j h_j Z_j + \Gamma \sum_j X_j  \, .
\eeq
for $\Gamma > 0$.  A slightly less trivial example is the two-body model
\beq\label{eq:Hising2}
H =\sum_{i,j} J_{ij} Z_i Z_j + \Gamma \sum_{\langle i,j\rangle} X_i X_j  \, ,
\eeq
provided that the underlying connectivity $\langle i,j\rangle$ of the two-body $X$ terms is bi-partite (allowing only even cycles).

It is also interesting to note that any single-qubit Hamiltonian is necessarily also sign-problem-free. In this case, the Hamiltonian is  $H=D_0 P_0+ D_1 P_1$ as described in Sec.~\ref{sec:single_spin_ex}. Since here the $S_{{\bf{i}}_q}$ are sequences consisting of only one type of non-identity permutation matrices, namely $P_1=X$, the expansion order $q$ must be even for $S_{{\bf{i}}_q}$ to evaluate to the identity element. This in turn results in $\left[\prod_{j=1}^q (-d^{(i_j)}_{z_j})\right] = {(\alpha^2_1+\alpha^2_2)}^{q/2}$ being strictly nonnegative. The same is however not true for a single qutrit in which case a sign problem may arise. 

\section{The QMC algorithm \label{sec:QMCalgo}}

Having derived the series expansion of the partition function for permutation-represented Hamiltonians, we are now in a position to discuss a QMC algorithm that can be associated with the above expansion. 

\subsection{QMC configurations and GBW calculation}

As was discussed above, a configuration \hbox{${\cal C}=\{|z\rangle, S_{{\bf{i}}_q}\}$} is a pair of 
a classical state and a product $S_{{\bf{i}}_q}$ of permutation operators that must evaluate to the identity element $P_0=\mathbb{1}$. To take full advantage of our partition function decomposition above, we treat the off-diagonal permutation terms $\{ P_j\}$ in our QMC algorithm as elements in a permutation group $G$ (with matrix product as the group operation). Since the elements $\{ P_j \}$ appearing in the Hamiltonian may not form a complete group, we shall treat any additional element $P_{j'}$ required to complete the set to form a group as appearing in the Hamiltonian with an associated diagonal matrix $D_{j'}=0$ [see Eq.~\eqref{eq:basic}]. 

The pair ${\cal C}$ induces a list of states \hbox{$\{|z_0\rangle =|z\rangle ,|z_1\rangle ,\ldots,|z_q\rangle =|z\rangle \}$}, which in turn also generates a corresponding {multiset} of diagonal 
energies $E_{{\cal C}}=\{E_{z_0},E_{z_1},\ldots,E_{z_q}\}$ of not-necessarily-distinct values (recall that $E_{z_i} = \langle z_i | H_c | z_i\rangle$). 
For systems with discrete energy values, the multiset can be stored efficiently in a `multiplicity table' 
$M_{{\cal C}}=\{ m_0,m_1,\ldots, m_j,\ldots\}$, where $m_j$ is the multiplicity of the energy $E_{z_j}$ in the multiset.

Given the multiset $E_{{\cal C}}$, the evaluation of the GBW $W_{{\cal C}}$ follows from its definition as a function of divided differences (the reader is referred to Ref.~\cite{ODE} for a more detailed description). 
The calculation of a GBW consisting of $q$ permutation operators requires the evaluation of a divided-differences exponential with $(q+1)$ energies. This calculation can be accomplished with at most $O(q)$ operations~\cite{Zivcovich2019,effDivDiff}.

\subsection{Initial state}
At this point we can consider a QMC algorithm based on the partition function expansion generating the weights 
$W_{{\cal C}}$, Eq.~(\ref{eq:gbw}). The Markov process would start with the initial configuration \hbox{$\mathcal{C}_0=\{|z\rangle, S_0=\mathbb{1}\}$} where $|z\rangle$ is a randomly generated initial classical state.
The weight of this initial configuration is 
\beq
W_{{\cal C}_0}=\e^{-\beta [E_z]}=\e^{-\beta E_z} \,,
\eeq
i.e., the classical Boltzmann weight of the initial random state $|z\rangle$. 

\subsection{Updates}

\begin{figure*}[htp]
\includegraphics[width=.9\textwidth]{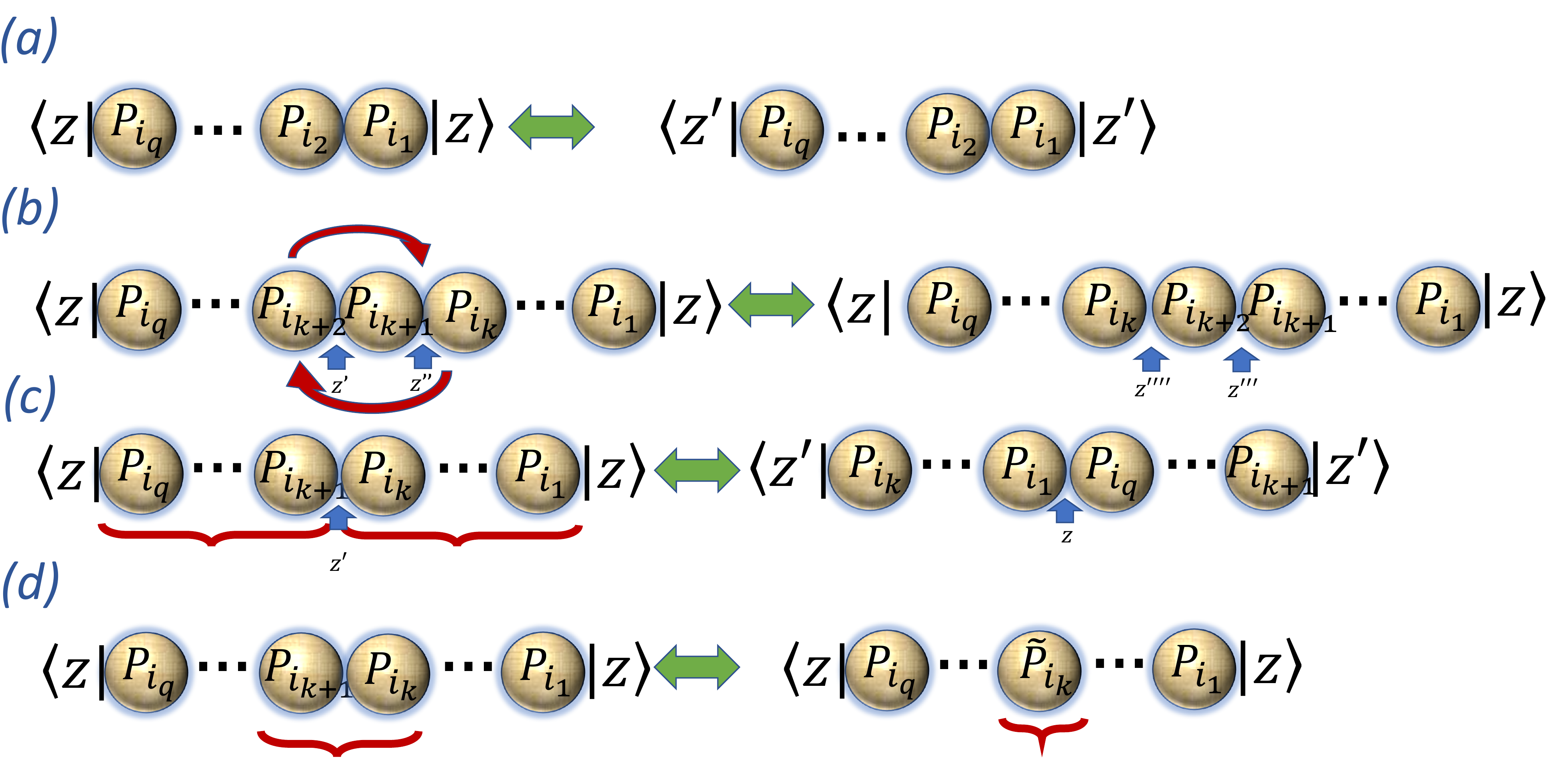}
\caption{\label{fig:updates} Basic update moves of the QMC algorithm. (a) Classical moves (e.g., a single bit flip), whereby only the initial state $z$ is changed to $z'$ leaving $S_{{\bf{i}}_q}$ unchanged. 
(b) Cyclic rotation, whereby two adjacent sequences of group elements (in this case $P_{i_k}$ and $P_{i_{k+1}}P_{i_{k+2}}$) whose product is the identity operation are interchanged, changing their internal classical states. (c) Block swap, whereby two partitions of the sequence $S_{{\bf{i}}_q}$ are interchanged.  This also changes the initial state from $z$ to $z'$ as well as the ordering of $S_{{\bf{i}}_q}$. (d) Cycle completion, whereby a sub-sequence of operators is replaced by an equivalent one (in this case, $P_{i_k} P_{i_{k+1}}$ is replaced by $\tilde{P}_{i_k}$. This is the only update where the number of group element (equivalently, the expansion order of the configuration) may change.
}
\end{figure*}

We next describe the basic update moves for the algorithm. These are also succinctly summarized in Fig.~\ref{fig:updates}.

\subsubsection{Classical moves}
Classical moves are any moves that involve a manipulation of the classical state $|z\rangle$ while leaving $S_{{\bf{i}}_q}$ unchanged [see Fig.~\ref{fig:updates}(a)]. 
In a single bit-flip classical move, a spin from the classical bit-string state $\ket{z}$ of ${\cal C}$ is picked randomly and is flipped, generating a state $\ket{z'}$ and hence a new configuration ${\cal C}'$.  Calculating the weight of  ${\cal C}'$ requires recalculating the energies associated with the product $S_{{\bf{i}}_q}$ leading to a new energy multiset $E_{{\cal C}'}$ and can become computationally intensive if $q$ is large. Classical moves should  therefore be attempted with low probabilities if $q$ is large. Simply enough, the acceptance probability for a classical move is
\beq\label{eq:Pmet}
p = \min \left( 1,\frac{W_{{\cal C}'}}{W_{{\cal C}}} \right)=\min \left( 1,\frac{\e^{-\beta [E_{{\cal C}'}]}}{\e^{-\beta [E_{{\cal C}}]}} \right)\,,
\eeq
where $\e^{-\beta [E_{{\cal C}}]}$ is a shorthand for $\e^{-\beta [E_{z_0},E_{z_1},\ldots,E_{z_q}]}$ of configuration ${\cal C}$ and likewise for ${\cal C}'$.

In the absence of a quantum part to the Hamiltonian ($D_j=0$ for all $j>0$), not only are classical moves the only moves necessary, they are also the only moves that have nonzero acceptance probabilities. Since the initial configuration of the QMC algorithm is a random classical configuration $|z\rangle$ and an empty operator sequence $S_0=\mathbb{1}$, for a purely classical Hamiltonian, the algorithm automatically reduces to a classical thermal algorithm keeping the size of the imaginary-time dimension at zero ($q=0$) for the duration of the simulation. 

 \subsubsection{Cyclic rotations}
The `cyclic rotation' move, [Fig.~\ref{fig:updates}(b)], consists of identifying short sub-sequences, or cycles, of consecutive operators in the sequence $S_{{\bf{i}}_q}$, whose product is the identity element, i.e.,
sub-sequences that obey
\beq
P_{i_j} \cdots P_{i_{j+C}}=\mathbb{1} \,.
\eeq 
Depending on the nature of the operators, preparing a lookup table of short cycles that evaluate to the identity may prove useful.
Once a cycle is identified, a random cycle rotation is attempted. Here, a random internal insertion point within the sub-sequence is picked and a rotation is attempted:
\beq
P_{i_j} \cdots P_{i_{k}} P_{i_{k+1}}\cdots P_{i_{j+C}} \to
P_{i_{k+1}}\cdots P_{i_{j+C}} P_{i_j} \cdots P_{i_{k}} \,.
\eeq
The rotated sequence also evaluates to the identity. 
Since the internal classical states between the elements in the cycle may change by the rotation,  the rotation involves adding new energies $ \{E_{z'}\ldots\}$ and removing old ones $\{E_{z'}\ldots\}-\{E_{z}\ldots\}$ from the energy multiset. Short cycles should therefore be preferred.  The acceptance probability for the move is as in Eq.~(\ref{eq:Pmet}) with $E_{{\cal C}'}  = E_{{\cal C}} + \{E_{z'}\ldots\}-\{E_{z}\ldots\}$. 

\subsubsection{Block-swap} \label{sec:blockswap}
A block swap  [Fig.~\ref{fig:updates}(c)] is an update that involves a change of the classical state $z$. Here, a random position $k$ in the product $S_{{\bf{i}}_q}$ is picked such that the product is split into two (non-empty) sub-sequences, $S_{{\bf{i}}_q}=S_2 S_1$, with $S_1 = P_{i_k} \cdots P_{i_{1}}$ and $S_2 = P_{i_{q}} \cdots P_{i_{k+1}}$.  The classical state $\ket{z'}$ at position $k$ in the product is given by
\beq
|z'\rangle=S_1|z\rangle=P_{i_k} \cdots P_{i_1}|z\rangle \,,
\eeq
where $\ket{z}$ is the classical state of the current configuration.  The state $\ket{z'}$ has energy $E_{z'}$, and the state $\ket{z}$ has energy $E_{z}$.  The new block-swapped configuration is ${\cal C}'=\{|z'\rangle, S_1 S_2\}$.
The multiplicity table of this configuration differs from that of the current configuration by having one fewer $E_{z}$ state and one additional $E_{z'}$ state.  The weight of the new configuration is then proportional to $e^{-\beta [E_{{\cal C}'}]}$
where the multiset $E_{{\cal C}'}=E_{{\cal C}} + \{E_{z'}\} - \{E_{z}\}$. 
The acceptance probability is as in Eq.~(\ref{eq:Pmet}) with the aforementioned $E_{{\cal C}'}$.

\subsubsection{Cycle completion}
The moves presented so far have left the number of group elements in the sequence, or expansion order, namely $q$, unchanged. 
The cycle completion move has the effect of changing the value of $q$.  
A lookup table of short cycles obeying
\beq
P_{i_j} \cdots P_{i_{j+C}}=\mathbb{1}
\eeq 
will be helpful in this case. 
The cycle completion move identifies a sub-cycle in the sequence $S_{{\bf{i}}_q}$, e.g., 
$
P_{i_j} \cdots P_{i_{k}} 
$
and replaces it with its complement 
\beq
\left(P_{i_{k+1}} \cdots P_{i_{j+C}} \right)^{-1} = P_{i_{j+C}}^{-1} \cdots P_{i_{k+1}}^{-1} \,.
\eeq
Note that the inverses of permutation matrices are also permutation matrices and are therefore also present in $G$. 

For concreteness, let us consider the case of subsequences of length two.  We randomly pick a point $k \in [0,q]$ in the sequence.  With probability $1/4$, the subsequence is taken to be $P_0 P_0$, $P_0 P_{i_k}$, $P_{i_{k-1}} P_0$, $P_{i_{k-1}} P_{i_k}$.\footnote{Note that if $k = 0$ or $q$, namely, the edges of the sequence, then two of the choices correspond to non-starters as there is an operator only to one side of the insertion point and so either $P_{i_{k-1}}$ or $P_{i_k}$ are not defined. }  The identified subsequence is replaced by its complement, resulting in a new configuration $\mathcal{C}'$. Because we can interpret  $P_0^{-1} = P_0 = P_j P_j^{-1}$ (for any arbitrary index $j$) and so on, the cycle completion move can grow and shrink the sequence.  The acceptance probability is as in Eq.~(\ref{eq:Pmet}) with the new configuration.
 
\subsection{Measurements}

Having reviewed the various update moves we next turn to discuss measurements within the algorithm.  
\subsubsection{Diagonal measurements}
A diagonal operator $\Lambda$ obeys $\Lambda|z\rangle=\lambda(z)|z\rangle$ where $\lambda(z)$ is a number that depends both on the operator and the state it acts on. Since $\langle z| \Lambda S_{{\bf{i}}_q} |z\rangle=\lambda(z)\langle z| S_{{\bf{i}}_q}|z\rangle$, for any given configuration \hbox{${\cal C}=(|z\rangle,S_{{\bf{i}}_q})$}, there is a contribution $\lambda=\lambda(z)$ to the diagonal operator thermal average $\langle \Lambda \rangle$. To improve statistics, one may also consider rotations in (the periodic) imaginary time. To do that, we may consider `virtual' block-swap moves (see Sec.~\ref{sec:blockswap}) that rotate $S_{{\bf{i}}_q}$ and as a result also change the classical configuration from $\ket{z}$ to $\ket{z_i}$.
The contribution to the expectation value of a diagonal operator $\Lambda$ thus becomes:
\beq
\lambda=\frac1{{\cal Z}} \sum_{i=0}^{q-1} \lambda(z_i) \e^{-\beta [E_{{\cal C}_i}]} \ .
\eeq
where  $E_{{\cal C}_i}$ is the energy multiset associated with configuration ${\cal C}_i$ whose multiset is
$E_{{\cal C}_i} =E_{{\cal C}} + \{E_{z_i}\} - \{ E_{z}\}$ (recall that $z_0 \equiv z$, so $E_{{\cal C}_0} = E_{{\cal C}}$). The normalization factor ${\cal Z}$ above is the sum
\beq
{\cal Z}=\sum_{j=0}^{q-1} \e^{-\beta [E_{{\cal C}_j}]} =\sum_j m_j \e^{-\beta [E_{{\cal C}_j}]}
\eeq
over all nonzero multiplicities $m_j$. 
In the case where $\Lambda=H_\D$ the above expression simplifies to:
\beq
\lambda = \frac1{{\cal Z}}\sum_{i=0}^{q-1} E_{z_i} \e^{-\beta [E_{{\cal C}_i}]}=\frac1{{\cal Z}}\sum_{j} m_j E_{z_j}  \e^{-\beta [E_{{\cal C}_j}]}  \,.
\eeq
\subsubsection{Off-diagonal measurements}
%
We next consider the case of measuring the expectation value of an off-diagonal operator $P_{k}$, namely, $\langle P_{k}\rangle$.  To do this, we interpret the instantaneous configuration as follows
\bea
W_{{\cal C}} &=& D_{(z,S_{{\bf{i}}_q})} \e^{-\beta [E_{{\cal C}}]} \bra{z}  S_{{\bf{i}}_q} \ket{z} =
 \left( \frac{d_{i_q} e^{-\beta [E_{\cal C}] }} {\e^{-\beta [E_{\cal C'}]}}\right) \nonumber\\
&\times& \left[D_{(z,S_{{\bf i}_{q-1}})}  \e^{-\beta [E_{{\cal C}'}]}   \bra{z}  S_{{\bf i}_{q-1}}P_{i_q} \ket{z}\right]  \,,
\eea
where ${\cal C}'$ is the configuration associated with the multiset $E_{{\cal C}'} = E_{{\cal C}}- \{ E_{z}\}$. 
In the above form, we can reinterpret the weight $W_{{\cal C}}$ as contributing 
\beq
p_{k}= \delta_{k,i_q} \frac{\e^{-\beta [E_{\cal C'}] }}{d^{(i_q)}_{z_q} e^{-\beta [E_{\cal C}]}} \,,
\eeq
to $\langle P_{k}\rangle$ where $d^{(i_q)}_{z_q}$ is the 'hopping strength' of $P_k$ Eq. (\ref{eq:hsAppC}).

As in the case of the diagonal measurements, one can take advantage of the periodicity in the imaginary time direction to improve statistics by rotating the sequence such that any of the elements of $S_{{\bf{i}}_q}$ becomes the last element of the sequence (see Sec.~\ref{sec:blockswap}), weighted accordingly by the block-swap probability.  By doing so, $P_{k}$ becomes
\bea
p_k &=& \sum_j \frac{\delta_{k,i_j}}{d^{(i_j)}_{z_j}} \frac{\e^{-\beta  E_{{\cal C}_j} } }{\sum_{j'=0}^{q-1} e^{-\beta [E_{{\cal C}_{j'}}] }} \frac{\e^{-\beta [E_{{\cal C}'}] }}{\e^{-\beta [E_{{\cal C}_j}]  }} \nonumber\\
&=& \frac{1}{{\cal Z}} \e^{-\beta [E_{{\cal C}'}]} \sum_j \frac{\delta_{k,i_j}}{d^{(i_j)}_{z_j}}\,,
\eea
where $E_{{\cal C}_i}  = E_{{\cal C}} + \{ E_{z_i}\}- \{ E_{z}\}$, the sum $\sum_j$ is over all rotated configurations ${\cal C}'$.
%
\subsubsection{Products of off-diagonal measurements}
%
The sampling of expectation values of the form $\langle P_{k_1} P_{k_2}\rangle$ proceeds very similarly to the single operator case except that now both operators must appear at the end of the sequence.  The argument proceeds similarly to the single off-diagonal measurement, and we have that  the contribution to the expectation value of $\langle P_{k_1} P_{k_2}\rangle $ is
\beq
p_{k_1,k_2} = \delta_{k_1,i_q} \delta_{k_2,i_{q-1}} \frac{ \e^{-\beta [E_{\cal C'}] }}{d^{(i_q)}_{z_q}d^{(i_{q-1})}_{z_{q-1}} e^{-\beta [E_{\cal C}]}}
\eeq
with $E_{{\cal C}'} = E_{{\cal C}} - \{E_{z},E_{z_{q-1}}\}$.  As in the single off-diagonal operator case, we can use the block-swap move to alter the elements at the end of the sequence, and for each pair of adjacent operators in the sequence obtain an improved contribution.  By doing so, $\langle P_{k_1} P_{k_2}\rangle $ becomes
\bea
P_{k_1,k_2} &=& \sum_j 
\frac{ \delta_{k_1,i_j} \delta_{k_2,i_{j-1}} }{d^{(i_j)}_{z_j}d^{(i_{j-1})}_{z_{j-1}} }
\frac{\e^{-\beta [E_{{\cal C}_j}] } }{\sum_{j'=0}^{q-1} e^{-\beta [E_{{\cal C}_{j'}}] }} \frac{\e^{-\beta  [E_{{\cal C}'_j}] }}{\e^{-\beta [E_{{\cal C}_j}]  }} \nonumber\\
&=&
\frac{1}{{\cal Z} } \sum_j \frac{ \delta_{k_1,i_j} \delta_{k_2,i_{j-1}} }{d^{(i_j)}_{z_j}d^{(i_{j-1})}_{z_{j-1}} } \e^{-\beta [E_{{\cal C}'_j}]}  \,,
\eea
where $E_{{\cal C}_k} =  E_{{\cal C}} + \{ E_{z_k}\}- \{ E_{z}\}$, $E_{{\cal C}'_i} = E_{{\cal C}} - \{E_{z},E_{z_{q-1}}\}$ with  $|z''\rangle = P_{k_2}|z'\rangle$ and $|z'\rangle$ is the classical state after the block swap. Similar to the single off-diagonal operator case, the sum $\sum_j$ is over all rotated configurations ${\cal C}'$ whose $S_{{\bf{i}}_q}$ ends with $P_{k_1} P_{k_2}$. 

Measurements of thermal averages of products of more than two off-diagonal operators can also be derived in a straightforward manner. 

\subsubsection{Improved measurements}

As will often happen, certain physical operators will have more than one representation as group element. 
E.g., if $P_3=P_1 P_2$, one could measure both the single operator $\langle P_3\rangle$ and the operator product 
$\langle P_1 P_2\rangle$ and combine the results.

\section{Results\label{sec:Results}}

In this section we present some results that highlight some of the advantages that PMR has to offer over existing methods. 
Specifically, we compare the performance of PMR over SSE on random 3-regular MAX2SAT instances augmented with a transverse field. This class of instances corresponds to a particular choice of the Ising Hamiltonian given in Eq.~\eqref{eq:Hising} whereby each spin is coupled antiferromagnetically (with strength $J_{ij} = 1$) with exactly three other spins picked at random. This class of instances is known to exhibit a quantum spin-glass phase transition and is notoriously difficult to simulate by standard QMC techniques, making it suitable to illustrate the strengths of the PMR algorithm (see Ref.~\cite{farhi:12} for more details).

In a subsequent section, we demonstrate the versatility of our formalism by considering the performance of PMR on a transverse-field Ising model augmented with random two-body $XX$ connections, which to the authors' knowledge cannot be readily implemented using existing methods~\cite{troyerXX}.

\subsection{PMR vs SSE: Transverse-field Ising model simulations\label{sec:bench}}
To demonstrate the advantages of PMR over existing state-of-the-art, we study random 3-regular MAX2SAT instances augmented with a transverse field. We study the thermal properties by utilizing a parallel tempering scheme~\cite{hukushima:96,marinari:98a,hen:11,farhi:12} for both PMR and SSE with 11 replicas at inverse temperatures, $\beta \in \left\{ 0.1, 0.2, 0.5, 1, 2, 5, 10, 20, 30, 40, 50 \right\}$.  We carry out simulations for 50 random MAX2SAT instances at sizes $N=96, 128$ and at transverse field strengths $\Gamma = 0.1, 0.4$. 

\begin{figure*}[htp]
\includegraphics[angle=0,scale=1,width=2.1\columnwidth]{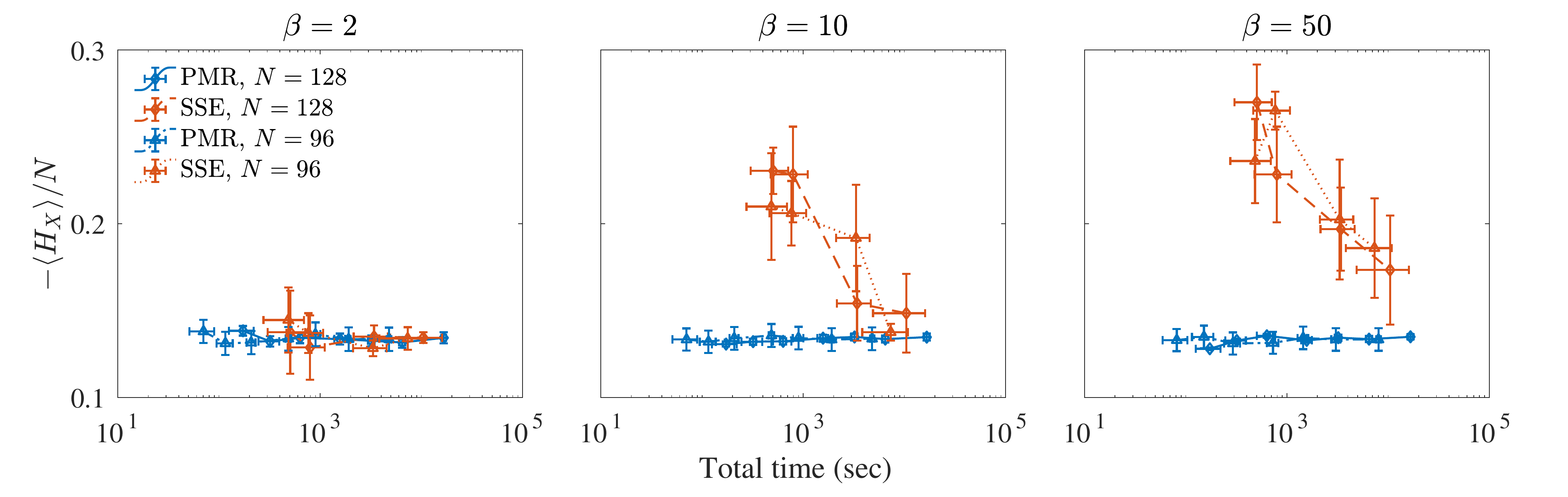}
\caption{{\bf Thermal average of the off-diagonal Hamiltonian as obtained by parallel tempering simulations using PMR (blue) and SSE (red) for $\Gamma=0.4$ as a function of simulation time.}  A subset of the results are shown for $\beta=2, 10$ and $50$. Each data point is the mean value over 50 random instances, with each simulation performing $500$ measurements of the $x$-magnetization-per-spin in the reported runtime (horizontal axis).  The error bars correspond to the $2\sigma$ confidence interval generated by $10^3$ bootstraps performed over the instances.}
\label{fig:BenchGamma04}
\end{figure*}

\begin{figure*}[htp]
\includegraphics[angle=0,scale=1,width=2.2\columnwidth]{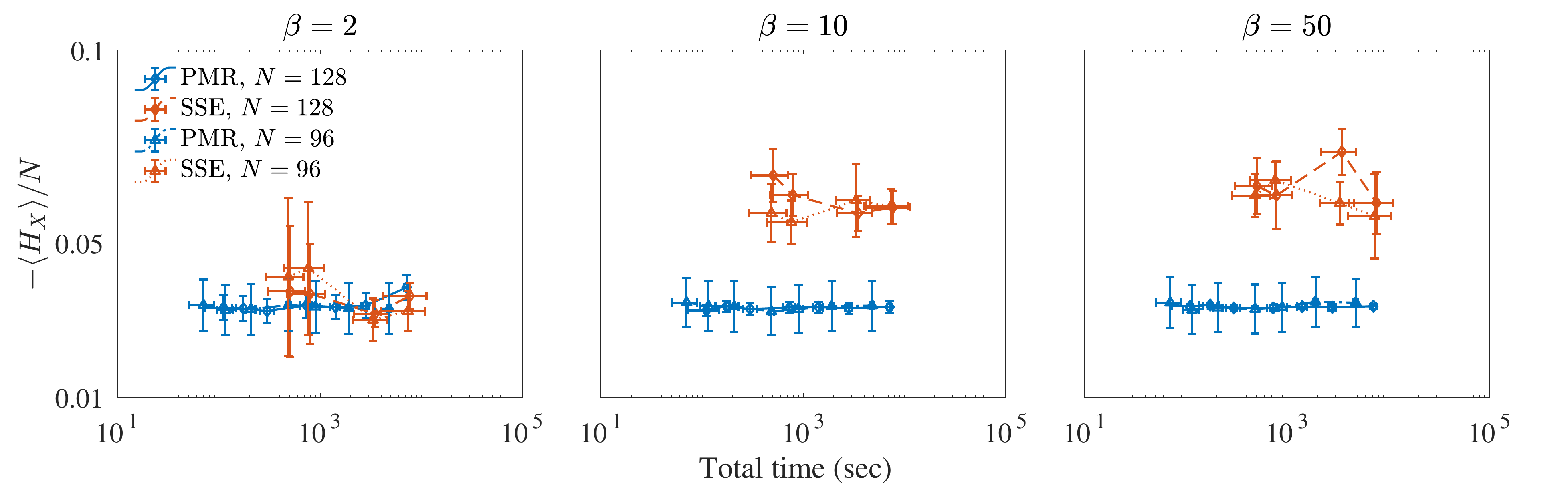}
\caption{{\bf Thermal average of the off-diagonal Hamiltonian as obtained by parallel tempering simulations using PMR (blue) and SSE (red) for $\Gamma=0.1$ as a function of simulation time.}  A subset of the results are shown for $\beta=2, 10$ and $50$. Each data point is the mean value over 50 random instances, with each simulation performing $500$ measurements of the $x$-magnetization-per-spin in the reported runtime (horizontal axis).  The error bars correspond to the $2\sigma$ confidence interval generated by $10^3$ bootstraps performed over the instances. }
\label{fig:BenchGamma01}
\end{figure*}

To quantify the performance of the algorithms, we fix the total number of measurements of our observables, and we vary the total number of updates between measurements.  Thus, we are able to measure the dependence of the estimate of the thermal expectation value with the total time spent de-correlating measurement samples.  The performance comparisons of the PMR algorithm against SSE are summarized in Figs.~\ref{fig:BenchGamma04} and~\ref{fig:BenchGamma01}.  Both figures depict the thermal average of $x$-magnetization as a function of simulation runtime (other observables exhibit similar behavior). 

As is evident from the figures, PMR is four or
more orders of magnitude faster than SSE, converging on average after only a few seconds in all cases.
On the other hand, for a large fraction of the instances, the SSE simulations did not finish
running over the 24 hour window allocated for each run. As expected, the difference in performance is even more pronounced in the more `classical' $\Gamma=0.1$ case (Fig.~\ref{fig:BenchGamma01}).

\subsection{Ising model with random $XX$ interactions}

We next illustrate the utility of our technique by studying a model that likely requires highly non-trivial implementations if studied by other QMC algorithms. 
We consider a transverse-field Ising model with random $XX$ interactions whose Hamiltonian is given by
\bea\label{eq:Isingeq}
H&=& s \sum_{\langle i j \rangle} Z_i Z_j -(1-s) \sum_i  X_i \nonumber\\
&-& b s( 1-s) \sum_{\langle i j \rangle} X_i X_j  \,.
\eea
Here, $s$ is a parameter in the range $(0,1)$, and $b\in \{0,1\}$ determines whether the two-body $X$ terms are absent ($b=0$) or present ($b=1$).
We examine underlying connectivity graphs that are Erd{\H o}s--R{\'e}nyi random, meaning we randomly pick a pair of spins to connect. The total number of edges for each instance is taken to be $n m /2$, where $m \in \left\{3,4,5 \right\}$ is the average degree of the graph (we focus only on single component graphs for simplicity). 

Hamiltonians of the above general form appear widely in the context of quantum annealing processes~\cite{kadowaki_quantum_1998}, where the system is evolved according to the above Hamiltonian by varying the parameter $s$ slowly in time from $s=0$ to $s=1$. The goal in quantum annealing is for the system to reach a state at the end of the anneal that has considerable overlap with the ground state manifold of the $Z$-dependent `problem' Hamiltonian, which in this case is a MaxCut instance (or a random antiferromagnetic)~\cite{farhi:12}. 
While in standard quantum annealing the two-body $X$ terms are normally absent (i.e., $b=0$), one is often interested in understanding the effects of augmenting the Hamiltonian with a `catalyst' --- an extra term that is hoped to reduce the amount of time required for the annealing process to take place (see, e.g., Refs.~\cite{durkin,2018arXiv181109980A,nonStoq3,PhysRevB.95.184416,10.3389/fict.2017.00002}). Setting $b=1$ can be viewed as an example of such a situation.

For $H$ in Eq.~\eqref{eq:Isingeq}, we have \hbox{$D_0=H_\cl=s \sum_{\langle i j \rangle} Z_i Z_j$} as well as one-body and two-body (in the $b=1$ case) off-diagonal $P_j$ operators: $\{X_i\} \cup \{X_i X_j\}_{\langle i j \rangle}$. The $D_j$ operators are all of the form $D_j =d_j \cdot \mathbb{1}$ where for the one body operators $d_j=-(1-s)$ and for the two-body operators $d_j=-s(1-s)$. That $d_j\leq0$ implies that the model is stoquastic and hence sign-problem-free. 
Given a configuration $\{ |z\rangle, S_q = P_{i_1} P_{i_2} \ldots P_{i_q} \}$, the QMC algorithm generated new configuration by either changing basis state $|z\rangle$ or subsequence of off-diagonal operator $S_q$. For our updates, we restrict to subsequences of length two which is enough to ensure ergodicity. The possible completion moves are summarized in Table~\ref{tab:moveXX}. For illustration, say $P_{i_1} = X_i X_k$ then new configuration  $\{ |z\rangle, S'_q = P'_{i_0} P'_{i_1} P_{i_2} \ldots P_{i_q} \}$ where $P'_{i_0} = X_iX_j$, $P'_{i_1} = X_jX_k$ can be generated by replacing $X_iX_k \rightarrow X_iX_j, X_j X_k$ and accepting the move with a probability satisfying detailed balance condition.


By inspecting the results of QMC simulations of the above Hamiltonian we are able to answer a number of questions that are relevant to quantum annealing. 
We first examine the variance of $H$ (denoted $\sigma^2_H$), which when close to 0 indicates that the thermal state is close to being purely in the ground state of the system (technically, any energy eigenstate of the system will give $\sigma^2_H = 0$).  For sufficiently low temperatures, this will always be the case, but the energy gap and the density  of states determine how low the temperature needs to be. 

\begin{figure}[htbp] 
   \centering
  \includegraphics[width=0.95\columnwidth]{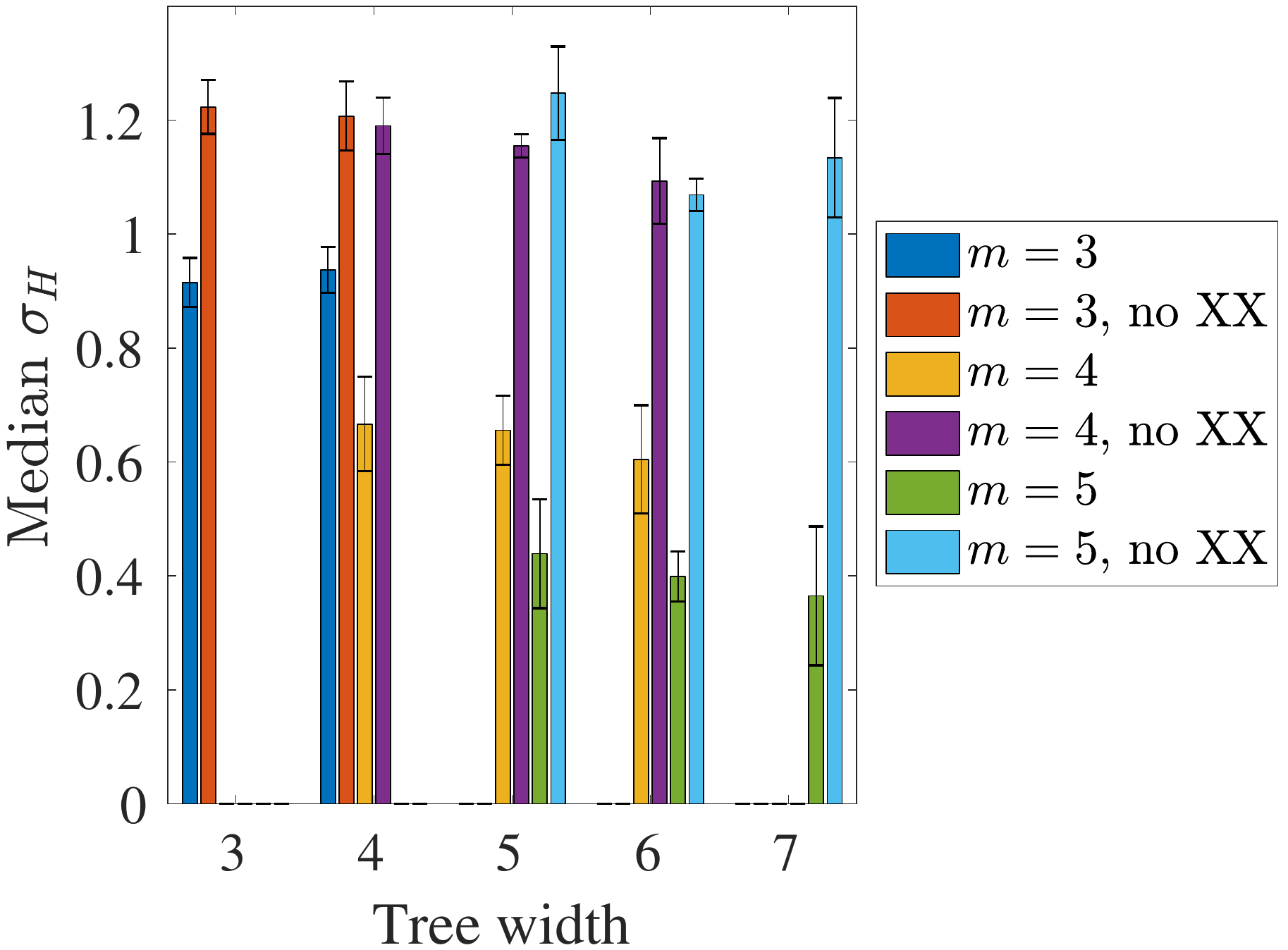} 
  \caption{Variance of $H$, denoted $\sigma^2_H$, as a function of the tree-width of the underlying graph (here, $n=16$, $s = 0.5$ and $\beta=2$). For $m=3,4,5$, we have 68, 89, and 99 instances.  The tree-width of each instance is identified, and each bar corresponds to the median value of $\sigma_H$ over the instances of a fixed $m$ and tree-width after $10^7$ QMC sweeps.  Error bars correspond to 95\% confidence interval calculated using a bootstrap over the instances of a fixed $m$ and tree-width.}
   \label{fig:IsingXX2}
\end{figure}

We therefore study the dependence of $\sigma^2_H$ on the instances' tree-widths.  This is shown in Fig.~\ref{fig:IsingXX2}. We see that while instances with different $m$ values may have the same tree-width, in the presence of $XX$ interactions there can be a significant difference in their $\sigma^2_H$ values. 
We make several observations.  First, we find that the Hamiltonian with $XX$ interactions requires more sweeps in order to de-correlate, i.e. thermalize.  Second, the differences for different $m$ are more substantial with $XX$ interactions than without them, indicating that the $XX$ interaction makes the spectrum much more susceptible to $m$. Third, larger $m$ values with $XX$ interactions tend to correspond to lower $\sigma^2_H$ values. This suggests that in the presence of $XX$ interactions, larger $m$ values are effectively `colder'.  Finally, we find that the tree-width makes little difference to the $\sigma^2_H$ values, with or without $XX$ interactions. 
 
\begin{table}
\begin{tabular}{|c|c c c |c |}
\hline
 &  &Move &  & Change in $q$ \\
\hline
i)&  $\mathbb{1}$& $\leftrightarrow $ &($X_i, X_i$) & $\pm 2$\\
 ii)&   $\mathbb{1}$ & $\leftrightarrow $ &($X_i X_j, X_i X_j$) & $ \pm 2$\\
 iii)&  $X_i X_j$ & $\leftrightarrow$ & ($X_i,X_j$) & $\pm1$ \\ 
iv)&     ($X_i, X_j$) & $\leftrightarrow$ & ($X_j,X_i$) & no change \\ 
v)&    ($X_i X_j, X_j X_k$) & $\leftrightarrow$ & $X_i X_k$ & $\pm 1$\\ 
\hline
\end{tabular}
\caption{\label{tab:moveXX} Cycle completion moves for the transverse field Ising model with two-body $X$ interactions. The moves include insertions or removals of pairs of identical one-body and two-body $X$ terms [i) and and ii)], the breaking up of a two-body term to its one-body constituents and the inverse operation [iii)], swapping [iv)] and the contraction of two operators into one [v)]. }
\end{table}

\begin{figure}[htbp] 
   \centering
   \includegraphics[width=0.95\columnwidth]{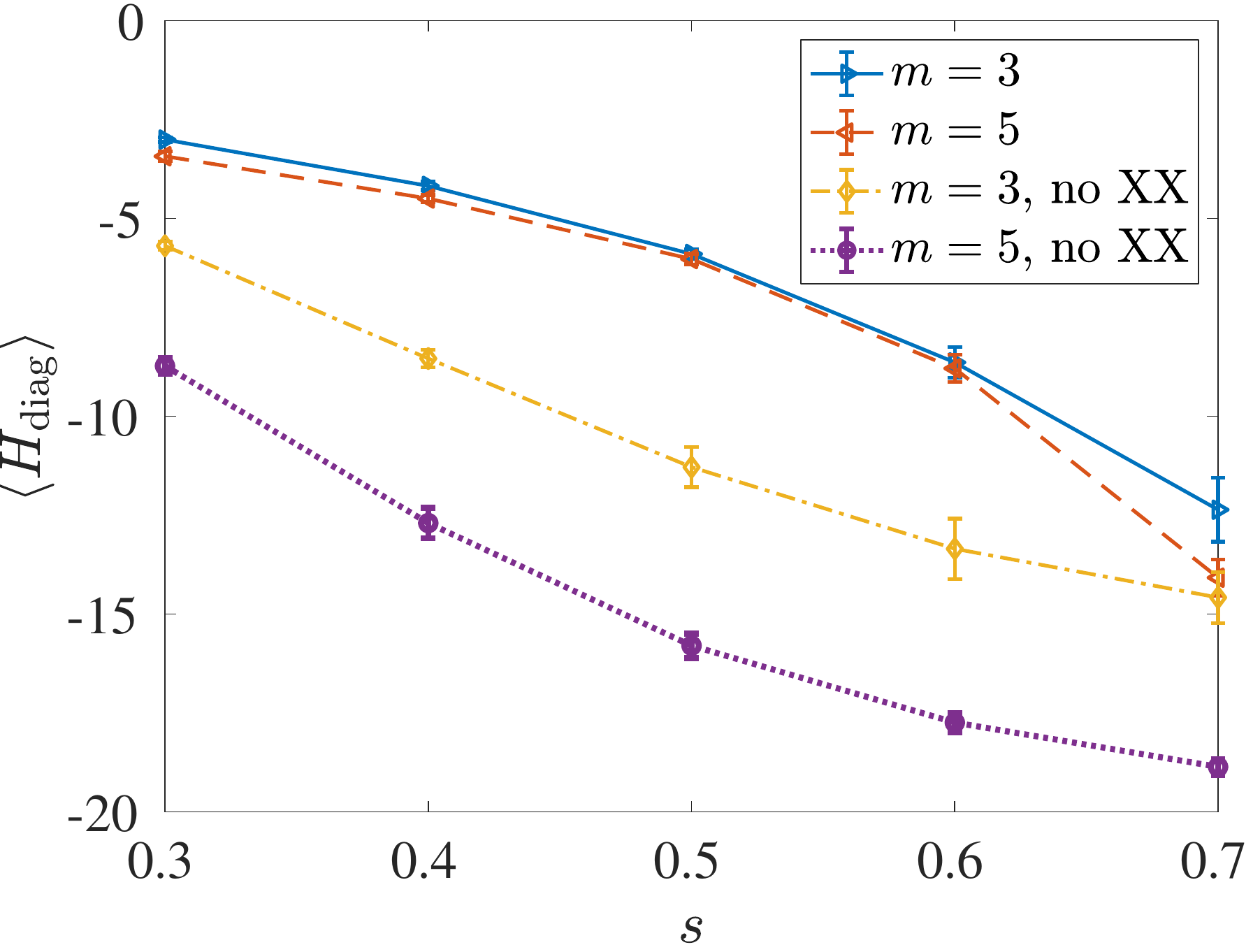} 
   \caption{Diagonal energy $\langle H_p\rangle$ as a function of $s$ with and without the $XX$ catalyst for $m=3$ and $m=5$ (here, $n=16, \beta=2$). Note that without the $XX$ catalyst the more connected graphs ($m=5$) have significantly lower diagonal energy than the $m=3$ case.}
   \label{fig:IsingXX3}
\end{figure}

Figure~\ref{fig:IsingXX3} shows the average diagonal energy as a function of the annealing parameter $s$ for the $b=0$ (no $XX$) and the $b=1$ case and for two different values of average graph degree, namely, $m=3$ and $m=5$. We find that the presence of the $XX$ catalyst has two important consequences: it minimizes the effect of the graph degree and significantly raises the average value.  The former effect suggests that the presence of an $XX$ catalyst will minimize performance differences in solving random MaxCut problems with different graph degrees. The latter effect is not surprising since the presence of both $X$ and $XX$ in the Hamiltonian means that we can expect the eigenstates to remain disordered for a larger region of $s$.

\section{Summary and discussion\label{sec:conclusions}}
We presented a parameter-free, Trotter-error free, universal quantum Monte Carlo scheme for the simulation of a broad range of physical models under a single unifying framework. Our technique enables the study of essentially any model on an equal footing.  In our approach, the quantum dimension consists of products of elements of permutation groups, allowing us to formulate update rules and measurement schemes independently of the model being studied.

We used our approach to clarify the emergence of the sign problem in the simulation of non-stoquastic physical models. In addition, studying the thermal properties of transverse-field Ising models, we illustrated the advantages of our technique over existing state-of-the-art, specifically the stochastic series expansion (SSE) algorithm. We showed that one of the features of the permutation matrix representation QMC that distinguishes it from SSE is that it bundles infinitely many SSE weights into a single weight. We demonstrated that this translates to orders-of-magnitude runtime advantages in most cases. 

We further illustrated the flexibility of our method by studying models with variable localities of interactions and underlying connectivities, namely, a transverse-field Ising model augmented with randomly placed two-body $X$ interactions, which to the authors' knowledge cannot be readily implemented using existing methods. For these, we found that the presence of the $XX$ interactions can mitigate differences associated with connectivity graphs of different degree. We expect that our algorithm can be further improved by implementing updates that utilize more `global' moves that, e.g., update longer cycles of operators. The extent to which this can translate to further runtime advantages remains an open question that we leave for future work.

We believe that the generality and flexibility of our algorithm will make it a useful tool in the study of physical models that have so far been inaccessible, cumbersome or too large to implement with existing techniques.

\begin{acknowledgments}
Computation for the work described here was supported by the University of Southern California's Center for High-Performance Computing (\url{http://hpcc.usc.edu}) and by ARO grant number W911NF1810227. 
IH is supported by the U.S. Department of Energy, Office of Science, Office of Advanced Scientific Computing Research (ASCR) Quantum Computing Application Teams (QCATS) program, under field work proposal number ERKJ347.
Work by LG was supported by the U.S. Department of Energy (DOE), Office of Science, Basic Energy Sciences (BES) under Award No. DE-SC0020280. 
Work by TA  is supported by the Office of the Director of National Intelligence (ODNI), Intelligence Advanced
Research Projects Activity (IARPA), via the U.S. Army Research Office
contract W911NF-17-C-0050. The U.S. Government is authorized to reproduce and distribute
reprints for Governmental purposes notwithstanding any copyright annotation thereon.
The views and conclusions contained herein are
those of the authors and should not be interpreted as necessarily
representing the official policies or endorsements, either expressed or
implied, of the ODNI, IARPA, or the U.S. Government.
\end{acknowledgments}
\pagebreak
\bibliography{refs}

\begin{thebibliography}{44}%
\makeatletter
\providecommand \@ifxundefined [1]{%
 \@ifx{#1\undefined}
}%
\providecommand \@ifnum [1]{%
 \ifnum #1\expandafter \@firstoftwo
 \else \expandafter \@secondoftwo
 \fi
}%
\providecommand \@ifx [1]{%
 \ifx #1\expandafter \@firstoftwo
 \else \expandafter \@secondoftwo
 \fi
}%
\providecommand \natexlab [1]{#1}%
\providecommand \enquote  [1]{``#1''}%
\providecommand \bibnamefont  [1]{#1}%
\providecommand \bibfnamefont [1]{#1}%
\providecommand \citenamefont [1]{#1}%
\providecommand \href@noop [0]{\@secondoftwo}%
\providecommand \href [0]{\begingroup \@sanitize@url \@href}%
\providecommand \@href[1]{\@@startlink{#1}\@@href}%
\providecommand \@@href[1]{\endgroup#1\@@endlink}%
\providecommand \@sanitize@url [0]{\catcode `\\12\catcode `\$12\catcode
  `\&12\catcode `\#12\catcode `\^12\catcode `\_12\catcode `\%12\relax}%
\providecommand \@@startlink[1]{}%
\providecommand \@@endlink[0]{}%
\providecommand \url  [0]{\begingroup\@sanitize@url \@url }%
\providecommand \@url [1]{\endgroup\@href {#1}{\urlprefix }}%
\providecommand \urlprefix  [0]{URL }%
\providecommand \Eprint [0]{\href }%
\providecommand \doibase [0]{http://dx.doi.org/}%
\providecommand \selectlanguage [0]{\@gobble}%
\providecommand \bibinfo  [0]{\@secondoftwo}%
\providecommand \bibfield  [0]{\@secondoftwo}%
\providecommand \translation [1]{[#1]}%
\providecommand \BibitemOpen [0]{}%
\providecommand \bibitemStop [0]{}%
\providecommand \bibitemNoStop [0]{.\EOS\space}%
\providecommand \EOS [0]{\spacefactor3000\relax}%
\providecommand \BibitemShut  [1]{\csname bibitem#1\endcsname}%
\let\auto@bib@innerbib\@empty
\bibitem [{\citenamefont {Landau}\ and\ \citenamefont
  {Binder}(2005)}]{Landau:2005:GMC:1051461}%
  \BibitemOpen
  \bibfield  {author} {\bibinfo {author} {\bibfnamefont {D.}~\bibnamefont
  {Landau}}\ and\ \bibinfo {author} {\bibfnamefont {K.}~\bibnamefont
  {Binder}},\ }\href@noop {} {\emph {\bibinfo {title} {A Guide to Monte Carlo
  Simulations in Statistical Physics}}}\ (\bibinfo  {publisher} {Cambridge
  University Press},\ \bibinfo {address} {New York, NY, USA},\ \bibinfo {year}
  {2005})\BibitemShut {NoStop}%
\bibitem [{\citenamefont {Barkema}(1999)}]{newman}%
  \BibitemOpen
  \bibfield  {author} {\bibinfo {author} {\bibfnamefont {M.~N. .~G.}\
  \bibnamefont {Barkema}},\ }\href@noop {} {\emph {\bibinfo {title} {Monte
  Carlo Methods in Statistical Physics}}}\ (\bibinfo  {publisher} {Oxford
  Uinversity Press},\ \bibinfo {year} {1999})\BibitemShut {NoStop}%
\bibitem [{\citenamefont {Aspuru-Guzik}\ \emph {et~al.}(2005)\citenamefont
  {Aspuru-Guzik}, \citenamefont {Dutoi}, \citenamefont {Love},\ and\
  \citenamefont {Head-Gordon}}]{Qchem1}%
  \BibitemOpen
  \bibfield  {author} {\bibinfo {author} {\bibfnamefont {A.}~\bibnamefont
  {Aspuru-Guzik}}, \bibinfo {author} {\bibfnamefont {A.~D.}\ \bibnamefont
  {Dutoi}}, \bibinfo {author} {\bibfnamefont {P.~J.}\ \bibnamefont {Love}}, \
  and\ \bibinfo {author} {\bibfnamefont {M.}~\bibnamefont {Head-Gordon}},\
  }\href {\doibase 10.1126/science.1113479} {\bibfield  {journal} {\bibinfo
  {journal} {Science}\ }\textbf {\bibinfo {volume} {309}},\ \bibinfo {pages}
  {1704} (\bibinfo {year} {2005})}\BibitemShut {NoStop}%
\bibitem [{\citenamefont {Kassal}\ \emph {et~al.}(2008)\citenamefont {Kassal},
  \citenamefont {Jordan}, \citenamefont {Love}, \citenamefont {Mohseni},\ and\
  \citenamefont {Aspuru-Guzik}}]{Qchem2}%
  \BibitemOpen
  \bibfield  {author} {\bibinfo {author} {\bibfnamefont {I.}~\bibnamefont
  {Kassal}}, \bibinfo {author} {\bibfnamefont {S.~P.}\ \bibnamefont {Jordan}},
  \bibinfo {author} {\bibfnamefont {P.~J.}\ \bibnamefont {Love}}, \bibinfo
  {author} {\bibfnamefont {M.}~\bibnamefont {Mohseni}}, \ and\ \bibinfo
  {author} {\bibfnamefont {A.}~\bibnamefont {Aspuru-Guzik}},\ }\href {\doibase
  10.1073/pnas.0808245105} {\bibfield  {journal} {\bibinfo  {journal}
  {Proceedings of the National Academy of Sciences}\ }\textbf {\bibinfo
  {volume} {105}},\ \bibinfo {pages} {18681} (\bibinfo {year}
  {2008})}\BibitemShut {NoStop}%
\bibitem [{\citenamefont {Lanyon}\ \emph {et~al.}(2010)\citenamefont {Lanyon},
  \citenamefont {Whitfield}, \citenamefont {Gillett}, \citenamefont {Goggin},
  \citenamefont {Almeida}, \citenamefont {Kassal}, \citenamefont {Biamonte},
  \citenamefont {Mohseni}, \citenamefont {Powell}, \citenamefont {Barbieri},
  \citenamefont {Aspuru-Guzik},\ and\ \citenamefont {White}}]{Qchem3}%
  \BibitemOpen
  \bibfield  {author} {\bibinfo {author} {\bibfnamefont {B.~P.}\ \bibnamefont
  {Lanyon}}, \bibinfo {author} {\bibfnamefont {J.~D.}\ \bibnamefont
  {Whitfield}}, \bibinfo {author} {\bibfnamefont {G.~G.}\ \bibnamefont
  {Gillett}}, \bibinfo {author} {\bibfnamefont {M.~E.}\ \bibnamefont {Goggin}},
  \bibinfo {author} {\bibfnamefont {M.~P.}\ \bibnamefont {Almeida}}, \bibinfo
  {author} {\bibfnamefont {I.}~\bibnamefont {Kassal}}, \bibinfo {author}
  {\bibfnamefont {J.~D.}\ \bibnamefont {Biamonte}}, \bibinfo {author}
  {\bibfnamefont {M.}~\bibnamefont {Mohseni}}, \bibinfo {author} {\bibfnamefont
  {B.~J.}\ \bibnamefont {Powell}}, \bibinfo {author} {\bibfnamefont
  {M.}~\bibnamefont {Barbieri}}, \bibinfo {author} {\bibfnamefont
  {A.}~\bibnamefont {Aspuru-Guzik}}, \ and\ \bibinfo {author} {\bibfnamefont
  {A.~G.}\ \bibnamefont {White}},\ }\href {http://dx.doi.org/10.1038/nchem.483}
  {\bibfield  {journal} {\bibinfo  {journal} {Nature Chemistry}\ }\textbf
  {\bibinfo {volume} {2}},\ \bibinfo {pages} {106 EP } (\bibinfo {year}
  {2010})}\BibitemShut {NoStop}%
\bibitem [{\citenamefont {Lonardoni}\ \emph {et~al.}(2014)\citenamefont
  {Lonardoni}, \citenamefont {Pederiva},\ and\ \citenamefont
  {Gandolfi}}]{1742-6596-529-1-012012}%
  \BibitemOpen
  \bibfield  {author} {\bibinfo {author} {\bibfnamefont {D.}~\bibnamefont
  {Lonardoni}}, \bibinfo {author} {\bibfnamefont {F.}~\bibnamefont {Pederiva}},
  \ and\ \bibinfo {author} {\bibfnamefont {S.}~\bibnamefont {Gandolfi}},\
  }\href {http://stacks.iop.org/1742-6596/529/i=1/a=012012} {\bibfield
  {journal} {\bibinfo  {journal} {Journal of Physics: Conference Series}\
  }\textbf {\bibinfo {volume} {529}},\ \bibinfo {pages} {012012} (\bibinfo
  {year} {2014})}\BibitemShut {NoStop}%
\bibitem [{\citenamefont {Chandrasekharan}\ and\ \citenamefont
  {Wiese}(1999)}]{PhysRevLett.83.3116}%
  \BibitemOpen
  \bibfield  {author} {\bibinfo {author} {\bibfnamefont {S.}~\bibnamefont
  {Chandrasekharan}}\ and\ \bibinfo {author} {\bibfnamefont {U.-J.}\
  \bibnamefont {Wiese}},\ }\href {\doibase 10.1103/PhysRevLett.83.3116}
  {\bibfield  {journal} {\bibinfo  {journal} {Phys. Rev. Lett.}\ }\textbf
  {\bibinfo {volume} {83}},\ \bibinfo {pages} {3116} (\bibinfo {year}
  {1999})}\BibitemShut {NoStop}%
\bibitem [{\citenamefont {Kieu}\ and\ \citenamefont
  {Griffin}(1994)}]{PhysRevE.49.3855}%
  \BibitemOpen
  \bibfield  {author} {\bibinfo {author} {\bibfnamefont {T.~D.}\ \bibnamefont
  {Kieu}}\ and\ \bibinfo {author} {\bibfnamefont {C.~J.}\ \bibnamefont
  {Griffin}},\ }\href {\doibase 10.1103/PhysRevE.49.3855} {\bibfield  {journal}
  {\bibinfo  {journal} {Phys. Rev. E}\ }\textbf {\bibinfo {volume} {49}},\
  \bibinfo {pages} {3855} (\bibinfo {year} {1994})}\BibitemShut {NoStop}%
\bibitem [{\citenamefont {Sandvik}(1992)}]{sandvik:92}%
  \BibitemOpen
  \bibfield  {author} {\bibinfo {author} {\bibfnamefont {A.~W.}\ \bibnamefont
  {Sandvik}},\ }\href@noop {} {\bibfield  {journal} {\bibinfo  {journal} {J.
  Phys, A}\ }\textbf {\bibinfo {volume} {25}},\ \bibinfo {pages} {3667}
  (\bibinfo {year} {1992})}\BibitemShut {NoStop}%
\bibitem [{\citenamefont {Sandvik}(1999)}]{sandvik:99}%
  \BibitemOpen
  \bibfield  {author} {\bibinfo {author} {\bibfnamefont {A.~W.}\ \bibnamefont
  {Sandvik}},\ }\href {\doibase 10.1103/PhysRevB.59.R14157} {\bibfield
  {journal} {\bibinfo  {journal} {Phys. Rev. B}\ }\textbf {\bibinfo {volume}
  {59}},\ \bibinfo {pages} {R14157} (\bibinfo {year} {1999})}\BibitemShut
  {NoStop}%
\bibitem [{\citenamefont {Prokof'ev}\ \emph {et~al.}(1998)\citenamefont
  {Prokof'ev}, \citenamefont {Svistunov},\ and\ \citenamefont
  {Tupitsyn}}]{prokofiev:98}%
  \BibitemOpen
  \bibfield  {author} {\bibinfo {author} {\bibfnamefont {N.~V.}\ \bibnamefont
  {Prokof'ev}}, \bibinfo {author} {\bibfnamefont {B.~V.}\ \bibnamefont
  {Svistunov}}, \ and\ \bibinfo {author} {\bibfnamefont {I.~S.}\ \bibnamefont
  {Tupitsyn}},\ }\href {\doibase 10.1134/1.558661} {\bibfield  {journal}
  {\bibinfo  {journal} {Journal of Experimental and Theoretical Physics}\
  }\textbf {\bibinfo {volume} {87}},\ \bibinfo {pages} {310} (\bibinfo {year}
  {1998})}\BibitemShut {NoStop}%
\bibitem [{\citenamefont {Mazzola}\ and\ \citenamefont
  {Troyer}(2017)}]{troyerXX}%
  \BibitemOpen
  \bibfield  {author} {\bibinfo {author} {\bibfnamefont {G.}~\bibnamefont
  {Mazzola}}\ and\ \bibinfo {author} {\bibfnamefont {M.}~\bibnamefont
  {Troyer}},\ }\href {http://stacks.iop.org/1742-5468/2017/i=5/a=053105}
  {\bibfield  {journal} {\bibinfo  {journal} {Journal of Statistical Mechanics:
  Theory and Experiment}\ }\textbf {\bibinfo {volume} {2017}},\ \bibinfo
  {pages} {053105} (\bibinfo {year} {2017})}\BibitemShut {NoStop}%
\bibitem [{\citenamefont {Albash}\ \emph {et~al.}(2017)\citenamefont {Albash},
  \citenamefont {Wagenbreth},\ and\ \citenamefont {Hen}}]{ODE}%
  \BibitemOpen
  \bibfield  {author} {\bibinfo {author} {\bibfnamefont {T.}~\bibnamefont
  {Albash}}, \bibinfo {author} {\bibfnamefont {G.}~\bibnamefont {Wagenbreth}},
  \ and\ \bibinfo {author} {\bibfnamefont {I.}~\bibnamefont {Hen}},\ }\href
  {\doibase 10.1103/PhysRevE.96.063309} {\bibfield  {journal} {\bibinfo
  {journal} {Phys. Rev. E}\ }\textbf {\bibinfo {volume} {96}},\ \bibinfo
  {pages} {063309} (\bibinfo {year} {2017})}\BibitemShut {NoStop}%
\bibitem [{\citenamefont {Hen}(2018)}]{ODE2}%
  \BibitemOpen
  \bibfield  {author} {\bibinfo {author} {\bibfnamefont {I.}~\bibnamefont
  {Hen}},\ }\href {http://stacks.iop.org/1742-5468/2018/i=5/a=053102}
  {\bibfield  {journal} {\bibinfo  {journal} {Journal of Statistical Mechanics:
  Theory and Experiment}\ }\textbf {\bibinfo {volume} {2018}},\ \bibinfo
  {pages} {053102} (\bibinfo {year} {2018})}\BibitemShut {NoStop}%
\bibitem [{\citenamefont {Handscomb}(1962)}]{handscomb1}%
  \BibitemOpen
  \bibfield  {author} {\bibinfo {author} {\bibfnamefont {D.~C.}\ \bibnamefont
  {Handscomb}},\ }\href@noop {} {\bibfield  {journal} {\bibinfo  {journal}
  {Proc. Camb. Phil. Soc.}\ }\textbf {\bibinfo {volume} {58}},\ \bibinfo
  {pages} {594–598} (\bibinfo {year} {1962})}\BibitemShut {NoStop}%
\bibitem [{\citenamefont {Handscomb}(1964)}]{handscomb2}%
  \BibitemOpen
  \bibfield  {author} {\bibinfo {author} {\bibfnamefont {D.~C.}\ \bibnamefont
  {Handscomb}},\ }\href@noop {} {\bibfield  {journal} {\bibinfo  {journal}
  {Proc. Camb. Phil. Soc.}\ }\textbf {\bibinfo {volume} {60}},\ \bibinfo
  {pages} {115—122} (\bibinfo {year} {1964})}\BibitemShut {NoStop}%
\bibitem [{\citenamefont {Sandvik}\ and\ \citenamefont
  {Kurkij\"arvi}(1991)}]{Sandvik1991}%
  \BibitemOpen
  \bibfield  {author} {\bibinfo {author} {\bibfnamefont {A.~W.}\ \bibnamefont
  {Sandvik}}\ and\ \bibinfo {author} {\bibfnamefont {J.}~\bibnamefont
  {Kurkij\"arvi}},\ }\href {\doibase 10.1103/PhysRevB.43.5950} {\bibfield
  {journal} {\bibinfo  {journal} {Phys. Rev. B}\ }\textbf {\bibinfo {volume}
  {43}},\ \bibinfo {pages} {5950} (\bibinfo {year} {1991})}\BibitemShut
  {NoStop}%
\bibitem [{\citenamefont {Sandvik}(2003)}]{sandvik:03}%
  \BibitemOpen
  \bibfield  {author} {\bibinfo {author} {\bibfnamefont {A.~W.}\ \bibnamefont
  {Sandvik}},\ }\href@noop {} {\bibfield  {journal} {\bibinfo  {journal} {Phys.
  Rev. E}\ }\textbf {\bibinfo {volume} {68}},\ \bibinfo {pages} {056701}
  (\bibinfo {year} {2003})}\BibitemShut {NoStop}%
\bibitem [{\citenamefont {Barker}(1979)}]{doi:10.1063/1.437829}%
  \BibitemOpen
  \bibfield  {author} {\bibinfo {author} {\bibfnamefont {J.~A.}\ \bibnamefont
  {Barker}},\ }\href {\doibase 10.1063/1.437829} {\bibfield  {journal}
  {\bibinfo  {journal} {The Journal of Chemical Physics}\ }\textbf {\bibinfo
  {volume} {70}},\ \bibinfo {pages} {2914} (\bibinfo {year} {1979})},\ \Eprint
  {http://arxiv.org/abs/https://doi.org/10.1063/1.437829}
  {https://doi.org/10.1063/1.437829} \BibitemShut {NoStop}%
\bibitem [{\citenamefont {Joyner}(2008)}]{gpm}%
  \BibitemOpen
  \bibfield  {author} {\bibinfo {author} {\bibfnamefont {D.}~\bibnamefont
  {Joyner}},\ }\href@noop {} {\emph {\bibinfo {title} {Adventures in group
  theory. Rubik's cube, Merlin's machine, and other mathematical toys}}}\
  (\bibinfo  {publisher} {Baltimore, MD: Johns Hopkins University Press},\
  \bibinfo {year} {2008})\BibitemShut {NoStop}%
\bibitem [{\citenamefont {Lewenstein}\ \emph {et~al.}(2012)\citenamefont
  {Lewenstein}, \citenamefont {Sanpera},\ and\ \citenamefont
  {Ahufinger}}]{lewenstein2012ultracold}%
  \BibitemOpen
  \bibfield  {author} {\bibinfo {author} {\bibfnamefont {M.}~\bibnamefont
  {Lewenstein}}, \bibinfo {author} {\bibfnamefont {A.}~\bibnamefont {Sanpera}},
  \ and\ \bibinfo {author} {\bibfnamefont {V.}~\bibnamefont {Ahufinger}},\
  }\href {https://books.google.com/books?id=WX4Xz7F6DdUC} {\emph {\bibinfo
  {title} {Ultracold Atoms in Optical Lattices: Simulating quantum many-body
  systems}}}\ (\bibinfo  {publisher} {OUP Oxford},\ \bibinfo {year}
  {2012})\BibitemShut {NoStop}%
\bibitem [{\citenamefont {Fisher}\ \emph {et~al.}(1989)\citenamefont {Fisher},
  \citenamefont {Weichman}, \citenamefont {Grinstein},\ and\ \citenamefont
  {Fisher}}]{PhysRevB.40.546_superfluid}%
  \BibitemOpen
  \bibfield  {author} {\bibinfo {author} {\bibfnamefont {M.~P.~A.}\
  \bibnamefont {Fisher}}, \bibinfo {author} {\bibfnamefont {P.~B.}\
  \bibnamefont {Weichman}}, \bibinfo {author} {\bibfnamefont {G.}~\bibnamefont
  {Grinstein}}, \ and\ \bibinfo {author} {\bibfnamefont {D.~S.}\ \bibnamefont
  {Fisher}},\ }\href {\doibase 10.1103/PhysRevB.40.546} {\bibfield  {journal}
  {\bibinfo  {journal} {Phys. Rev. B}\ }\textbf {\bibinfo {volume} {40}},\
  \bibinfo {pages} {546} (\bibinfo {year} {1989})}\BibitemShut {NoStop}%
\bibitem [{\citenamefont {Jaksch}\ and\ \citenamefont
  {Zoller}(2005)}]{JAKSCH200552_optical_lattice}%
  \BibitemOpen
  \bibfield  {author} {\bibinfo {author} {\bibfnamefont {D.}~\bibnamefont
  {Jaksch}}\ and\ \bibinfo {author} {\bibfnamefont {P.}~\bibnamefont
  {Zoller}},\ }\href {\doibase https://doi.org/10.1016/j.aop.2004.09.010}
  {\bibfield  {journal} {\bibinfo  {journal} {Annals of Physics}\ }\textbf
  {\bibinfo {volume} {315}},\ \bibinfo {pages} {52 } (\bibinfo {year}
  {2005})},\ \bibinfo {note} {special Issue}\BibitemShut {NoStop}%
\bibitem [{\citenamefont {{Giamarchi}}\ \emph {et~al.}(2008)\citenamefont
  {{Giamarchi}}, \citenamefont {{R{\"u}egg}},\ and\ \citenamefont
  {{Tchernyshyov}}}]{2008NatPh...4..198G_magnetic_insulator}%
  \BibitemOpen
  \bibfield  {author} {\bibinfo {author} {\bibfnamefont {T.}~\bibnamefont
  {{Giamarchi}}}, \bibinfo {author} {\bibfnamefont {C.}~\bibnamefont
  {{R{\"u}egg}}}, \ and\ \bibinfo {author} {\bibfnamefont {O.}~\bibnamefont
  {{Tchernyshyov}}},\ }\href {\doibase 10.1038/nphys893} {\bibfield  {journal}
  {\bibinfo  {journal} {Nature Physics}\ }\textbf {\bibinfo {volume} {4}},\
  \bibinfo {pages} {198} (\bibinfo {year} {2008})},\ \Eprint
  {http://arxiv.org/abs/0712.2250} {arXiv:0712.2250 [cond-mat.str-el]}
  \BibitemShut {NoStop}%
\bibitem [{\citenamefont {Hen}(2019)}]{signProbODE}%
  \BibitemOpen
  \bibfield  {author} {\bibinfo {author} {\bibfnamefont {I.}~\bibnamefont
  {Hen}},\ }\href {\doibase 10.1103/PhysRevE.99.033306} {\bibfield  {journal}
  {\bibinfo  {journal} {Phys. Rev. E}\ }\textbf {\bibinfo {volume} {99}},\
  \bibinfo {pages} {033306} (\bibinfo {year} {2019})}\BibitemShut {NoStop}%
\bibitem [{\citenamefont {Whittaker}\ and\ \citenamefont
  {Robinson}(1967)}]{dd:67}%
  \BibitemOpen
  \bibfield  {author} {\bibinfo {author} {\bibfnamefont {E.~T.}\ \bibnamefont
  {Whittaker}}\ and\ \bibinfo {author} {\bibfnamefont {G.}~\bibnamefont
  {Robinson}},\ }in\ \href@noop {} {\emph {\bibinfo {booktitle} {The Calculus
  of Observations: A Treatise on Numerical Mathematics}}}\ (\bibinfo
  {publisher} {New York: Dover},\ \bibinfo {address} {New York},\ \bibinfo
  {year} {1967})\BibitemShut {NoStop}%
\bibitem [{\citenamefont {de~Boor}(2005)}]{deboor:05}%
  \BibitemOpen
  \bibfield  {author} {\bibinfo {author} {\bibfnamefont {C.}~\bibnamefont
  {de~Boor}},\ }\href@noop {} {\bibfield  {journal} {\bibinfo  {journal}
  {Surveys in Approximation Theory}\ }\textbf {\bibinfo {volume} {1}},\
  \bibinfo {pages} {46} (\bibinfo {year} {2005})}\BibitemShut {NoStop}%
\bibitem [{\citenamefont {Bravyi}\ \emph {et~al.}(2008)\citenamefont {Bravyi},
  \citenamefont {DiVincenzo}, \citenamefont {Oliveira},\ and\ \citenamefont
  {Terhal}}]{Bravyi:QIC08}%
  \BibitemOpen
  \bibfield  {author} {\bibinfo {author} {\bibfnamefont {S.}~\bibnamefont
  {Bravyi}}, \bibinfo {author} {\bibfnamefont {D.~P.}\ \bibnamefont
  {DiVincenzo}}, \bibinfo {author} {\bibfnamefont {R.~I.}\ \bibnamefont
  {Oliveira}}, \ and\ \bibinfo {author} {\bibfnamefont {B.~M.}\ \bibnamefont
  {Terhal}},\ }\href {https://arxiv.org/abs/quant-ph/0606140} {\bibfield
  {journal} {\bibinfo  {journal} {Quant. Inf. Comp.}\ }\textbf {\bibinfo
  {volume} {8}},\ \bibinfo {pages} {0361} (\bibinfo {year} {2008})}\BibitemShut
  {NoStop}%
\bibitem [{\citenamefont {Bravyi}\ and\ \citenamefont
  {Hastings}(2014)}]{Bravyi:2014bf}%
  \BibitemOpen
  \bibfield  {author} {\bibinfo {author} {\bibfnamefont {S.}~\bibnamefont
  {Bravyi}}\ and\ \bibinfo {author} {\bibfnamefont {M.}~\bibnamefont
  {Hastings}},\ }\href {http://arXiv.org/abs/1410.0703} {\bibfield  {journal}
  {\bibinfo  {journal} {arXiv:1410.0703}\ } (\bibinfo {year}
  {2014})}\BibitemShut {NoStop}%
\bibitem [{\citenamefont {Marvian}\ \emph {et~al.}(2019)\citenamefont
  {Marvian}, \citenamefont {Lidar},\ and\ \citenamefont
  {Hen}}]{marvianLidarHen}%
  \BibitemOpen
  \bibfield  {author} {\bibinfo {author} {\bibfnamefont {M.}~\bibnamefont
  {Marvian}}, \bibinfo {author} {\bibfnamefont {D.~A.}\ \bibnamefont {Lidar}},
  \ and\ \bibinfo {author} {\bibfnamefont {I.}~\bibnamefont {Hen}},\ }\href
  {\doibase 10.1038/s41467-019-09501-6} {\bibfield  {journal} {\bibinfo
  {journal} {Nature Communications}\ }\textbf {\bibinfo {volume} {10}},\
  \bibinfo {pages} {1571} (\bibinfo {year} {2019})}\BibitemShut {NoStop}%
\bibitem [{\citenamefont {Gupta}\ and\ \citenamefont
  {Hen}(2019)}]{elucidating}%
  \BibitemOpen
  \bibfield  {author} {\bibinfo {author} {\bibfnamefont {L.}~\bibnamefont
  {Gupta}}\ and\ \bibinfo {author} {\bibfnamefont {I.}~\bibnamefont {Hen}},\
  }\href {\doibase 10.1002/qute.201900108} {\bibfield  {journal} {\bibinfo
  {journal} {Advanced Quantum Technologies}\ }\textbf {\bibinfo {volume} {2}},\
  \bibinfo {pages} {1900108} (\bibinfo {year} {2019})}\BibitemShut {NoStop}%
\bibitem [{\citenamefont {Zivcovich}(2019)}]{Zivcovich2019}%
  \BibitemOpen
  \bibfield  {author} {\bibinfo {author} {\bibfnamefont {F.}~\bibnamefont
  {Zivcovich}},\ }\href {http://drna.padovauniversitypress.it/2019/1/4}
  {\bibfield  {journal} {\bibinfo  {journal} {Dolomites Research Notes on
  Approximation}\ }\textbf {\bibinfo {volume} {12}},\ \bibinfo {pages} {28}
  (\bibinfo {year} {2019})}\BibitemShut {NoStop}%
\bibitem [{\citenamefont {Gupta}\ \emph {et~al.}(2020)\citenamefont {Gupta},
  \citenamefont {Barash},\ and\ \citenamefont {Hen}}]{effDivDiff}%
  \BibitemOpen
  \bibfield  {author} {\bibinfo {author} {\bibfnamefont {L.}~\bibnamefont
  {Gupta}}, \bibinfo {author} {\bibfnamefont {L.}~\bibnamefont {Barash}}, \
  and\ \bibinfo {author} {\bibfnamefont {I.}~\bibnamefont {Hen}},\ }\href
  {\doibase 10.1016/j.cpc.2020.107385} {\bibfield  {journal} {\bibinfo
  {journal} {Computer Physics Communications}\ }\textbf {\bibinfo {volume}
  {254}} (\bibinfo {year} {2020}),\ 10.1016/j.cpc.2020.107385}\BibitemShut
  {NoStop}%
\bibitem [{\citenamefont {Farhi}\ \emph {et~al.}(2012)\citenamefont {Farhi},
  \citenamefont {Gosset}, \citenamefont {Hen}, \citenamefont {Sandvik},
  \citenamefont {Shor}, \citenamefont {Young},\ and\ \citenamefont
  {Zamponi}}]{farhi:12}%
  \BibitemOpen
  \bibfield  {author} {\bibinfo {author} {\bibfnamefont {E.}~\bibnamefont
  {Farhi}}, \bibinfo {author} {\bibfnamefont {D.}~\bibnamefont {Gosset}},
  \bibinfo {author} {\bibfnamefont {I.}~\bibnamefont {Hen}}, \bibinfo {author}
  {\bibfnamefont {A.~W.}\ \bibnamefont {Sandvik}}, \bibinfo {author}
  {\bibfnamefont {P.}~\bibnamefont {Shor}}, \bibinfo {author} {\bibfnamefont
  {A.~P.}\ \bibnamefont {Young}}, \ and\ \bibinfo {author} {\bibfnamefont
  {F.}~\bibnamefont {Zamponi}},\ }\href {\doibase 10.1103/PhysRevA.86.052334}
  {\bibfield  {journal} {\bibinfo  {journal} {Phys. Rev. A}\ }\textbf {\bibinfo
  {volume} {86}},\ \bibinfo {pages} {052334} (\bibinfo {year}
  {2012})}\BibitemShut {NoStop}%
\bibitem [{\citenamefont {Hukushima}\ and\ \citenamefont
  {Nemoto}(1996)}]{hukushima:96}%
  \BibitemOpen
  \bibfield  {author} {\bibinfo {author} {\bibfnamefont {K.}~\bibnamefont
  {Hukushima}}\ and\ \bibinfo {author} {\bibfnamefont {K.}~\bibnamefont
  {Nemoto}},\ }\href {\doibase 10.1143/JPSJ.65.1604} {\bibfield  {journal}
  {\bibinfo  {journal} {J. Phys. Soc. Japan}\ }\textbf {\bibinfo {volume}
  {65}},\ \bibinfo {pages} {1604} (\bibinfo {year} {1996})},\ \Eprint
  {http://arxiv.org/abs/arXiv:cond-mat/9512035} {arXiv:cond-mat/9512035}
  \BibitemShut {NoStop}%
\bibitem [{\citenamefont {Marinari}\ \emph {et~al.}(1998)\citenamefont
  {Marinari}, \citenamefont {Parisi},\ and\ \citenamefont
  {Ruiz-Lorenzo}}]{marinari:98a}%
  \BibitemOpen
  \bibfield  {author} {\bibinfo {author} {\bibfnamefont {E.}~\bibnamefont
  {Marinari}}, \bibinfo {author} {\bibfnamefont {G.}~\bibnamefont {Parisi}}, \
  and\ \bibinfo {author} {\bibfnamefont {J.~J.}\ \bibnamefont {Ruiz-Lorenzo}},\
  }in\ \href@noop {} {\emph {\bibinfo {booktitle} {Spin Glasses and Random
  Fields}}},\ \bibinfo {editor} {edited by\ \bibinfo {editor} {\bibfnamefont
  {A.~P.}\ \bibnamefont {Young}}}\ (\bibinfo  {publisher} {World Scientific,
  Singapore},\ \bibinfo {address} {Singapore},\ \bibinfo {year} {1998})\
  p.~\bibinfo {pages} {59},\ \Eprint
  {http://arxiv.org/abs/(arXiv:cond-mat/9701016)} {(arXiv:cond-mat/9701016)}
  \BibitemShut {NoStop}%
\bibitem [{\citenamefont {Hen}\ and\ \citenamefont {Young}(2011)}]{hen:11}%
  \BibitemOpen
  \bibfield  {author} {\bibinfo {author} {\bibfnamefont {I.}~\bibnamefont
  {Hen}}\ and\ \bibinfo {author} {\bibfnamefont {A.~P.}\ \bibnamefont
  {Young}},\ }\href {\doibase 10.1103/PhysRevE.84.061152} {\bibfield  {journal}
  {\bibinfo  {journal} {Phys. Rev. E}\ }\textbf {\bibinfo {volume} {84}},\
  \bibinfo {pages} {061152} (\bibinfo {year} {2011})}\BibitemShut {NoStop}%
\bibitem [{\citenamefont {Kadowaki}\ and\ \citenamefont
  {Nishimori}(1998)}]{kadowaki_quantum_1998}%
  \BibitemOpen
  \bibfield  {author} {\bibinfo {author} {\bibfnamefont {T.}~\bibnamefont
  {Kadowaki}}\ and\ \bibinfo {author} {\bibfnamefont {H.}~\bibnamefont
  {Nishimori}},\ }\href {\doibase 10.1103/PhysRevE.58.5355} {\bibfield
  {journal} {\bibinfo  {journal} {Phys. Rev. E}\ }\textbf {\bibinfo {volume}
  {58}},\ \bibinfo {pages} {5355} (\bibinfo {year} {1998})}\BibitemShut
  {NoStop}%
\bibitem [{\citenamefont {Durkin}(2019)}]{durkin}%
  \BibitemOpen
  \bibfield  {author} {\bibinfo {author} {\bibfnamefont {G.~A.}\ \bibnamefont
  {Durkin}},\ }\href {\doibase 10.1103/PhysRevA.99.032315} {\bibfield
  {journal} {\bibinfo  {journal} {Phys. Rev. A}\ }\textbf {\bibinfo {volume}
  {99}},\ \bibinfo {pages} {032315} (\bibinfo {year} {2019})}\BibitemShut
  {NoStop}%
\bibitem [{\citenamefont {Albash}(2019)}]{2018arXiv181109980A}%
  \BibitemOpen
  \bibfield  {author} {\bibinfo {author} {\bibfnamefont {T.}~\bibnamefont
  {Albash}},\ }\href {\doibase 10.1103/PhysRevA.99.042334} {\bibfield
  {journal} {\bibinfo  {journal} {Phys. Rev. A}\ }\textbf {\bibinfo {volume}
  {99}},\ \bibinfo {pages} {042334} (\bibinfo {year} {2019})}\BibitemShut
  {NoStop}%
\bibitem [{\citenamefont {{Crosson}}\ \emph {et~al.}(2014)\citenamefont
  {{Crosson}}, \citenamefont {{Farhi}}, \citenamefont {{Yen-Yu Lin}},
  \citenamefont {{Lin}},\ and\ \citenamefont {{Shor}}}]{nonStoq3}%
  \BibitemOpen
  \bibfield  {author} {\bibinfo {author} {\bibfnamefont {E.}~\bibnamefont
  {{Crosson}}}, \bibinfo {author} {\bibfnamefont {E.}~\bibnamefont {{Farhi}}},
  \bibinfo {author} {\bibfnamefont {C.}~\bibnamefont {{Yen-Yu Lin}}}, \bibinfo
  {author} {\bibfnamefont {H.-H.}\ \bibnamefont {{Lin}}}, \ and\ \bibinfo
  {author} {\bibfnamefont {P.}~\bibnamefont {{Shor}}},\ }\href@noop {}
  {\bibfield  {journal} {\bibinfo  {journal} {ArXiv e-prints}\ } (\bibinfo
  {year} {2014})},\ \Eprint {http://arxiv.org/abs/1401.7320} {arXiv:1401.7320
  [quant-ph]} \BibitemShut {NoStop}%
\bibitem [{\citenamefont {Hormozi}\ \emph {et~al.}(2017)\citenamefont
  {Hormozi}, \citenamefont {Brown}, \citenamefont {Carleo},\ and\ \citenamefont
  {Troyer}}]{PhysRevB.95.184416}%
  \BibitemOpen
  \bibfield  {author} {\bibinfo {author} {\bibfnamefont {L.}~\bibnamefont
  {Hormozi}}, \bibinfo {author} {\bibfnamefont {E.~W.}\ \bibnamefont {Brown}},
  \bibinfo {author} {\bibfnamefont {G.}~\bibnamefont {Carleo}}, \ and\ \bibinfo
  {author} {\bibfnamefont {M.}~\bibnamefont {Troyer}},\ }\href {\doibase
  10.1103/PhysRevB.95.184416} {\bibfield  {journal} {\bibinfo  {journal} {Phys.
  Rev. B}\ }\textbf {\bibinfo {volume} {95}},\ \bibinfo {pages} {184416}
  (\bibinfo {year} {2017})}\BibitemShut {NoStop}%
\bibitem [{\citenamefont {Nishimori}\ and\ \citenamefont
  {Takada}(2017)}]{10.3389/fict.2017.00002}%
  \BibitemOpen
  \bibfield  {author} {\bibinfo {author} {\bibfnamefont {H.}~\bibnamefont
  {Nishimori}}\ and\ \bibinfo {author} {\bibfnamefont {K.}~\bibnamefont
  {Takada}},\ }\href {\doibase 10.3389/fict.2017.00002} {\bibfield  {journal}
  {\bibinfo  {journal} {Frontiers in ICT}\ }\textbf {\bibinfo {volume} {4}},\
  \bibinfo {pages} {2} (\bibinfo {year} {2017})}\BibitemShut {NoStop}%
\bibitem [{\citenamefont {Rotman}(1995)}]{groupTheory}%
  \BibitemOpen
  \bibfield  {author} {\bibinfo {author} {\bibnamefont {Rotman}},\ }\href@noop
  {} {\emph {\bibinfo {title} {An Introduction to the Theory of Groups}}}\
  (\bibinfo  {publisher} {New York: Springer-Verlag},\ \bibinfo {year}
  {1995})\BibitemShut {NoStop}%
\end{thebibliography}%

\appendix

\section{Finite dimensional permutation matrix representations\label{app:PermutationM}}
To show that any finite-dimensional matrix can be written in the form of Eq.~\eqref{eq:basic}, we will make use of `cycle notation'~\cite{groupTheory} --- a compact representation of permutations --- to represent the permutation matrices $P_j$. We start with some terminology. 

A cycle is a string of integers that represents an element of the symmetric permutation group $S_n$, which cyclically permutes these integers and fixes all other integers. For example cycle $(a_1, a_2, \ldots , a_m)$ is the permutation that sends $a_i$ to $a_{i+1}$, $1\leq i \leq m-1$ and sends $a_m$ to $a_1$. The cycle given in the above example is an $m$-cycle. In general, any element $\sigma \in S_n$ can be written as a product of $k$ cycles as 
$(a_1 \; a_2 \ldots a_{m_1})(a_{m_1 + 1} \; a_{m_1 + 2} \ldots a_{m_2} ) \ldots$ $(a_{m_{k-1}+1} \; a_{m_{k-1}+2} \ldots a_{m_k}) $. The order of a permutation $\sigma$ is defined as the smallest positive integer $p$ such that $\sigma^{p}$ is the identity element. In this notation, the action of $\sigma$ on any number from $1$ to $n$ can be determined as follows.  If $a$ appears at the right end of one of the $k$ cycles, then $\sigma(a)$ is the integer at the start of the cycle to which $a$ belongs. If an integer $a$ does not appear at the right end of one of the $k$ cycles, then $\sigma(a)$ is the integer to the right of $a$ in the cycle to which $a$ belongs.\\

For concreteness, let us write the $3 \times 3$ permutation matrices used in Sec.~\ref{sec:qutrit} in cycle notation.
\begin{equation}
P_1 = P =
\begin{bmatrix}
0 & 0 & 1 \\
1 & 0 & 0 \\
0 & 1 & 0 \\
\end{bmatrix}
\equiv (1,2,3) \,,
\end{equation}
\begin{equation}
P_2 = P^2 = 
\begin{bmatrix}
0 & 1 & 0 \\
0 & 0 & 1 \\
1 & 0 & 0 \\
\end{bmatrix}
\equiv (1,3,2) \,.
\end{equation}
\\
The identity operation can thus be written as 
\begin{equation}
P_0 = P^3 = \mathbb{1} =
\begin{bmatrix}
1 & 0 & 0 \\
0 & 1 & 0 \\
0 & 0 & 1 \\
\end{bmatrix}
\equiv (1) (2) (3) \,.
\end{equation}
\\
We illustrate the evaluation of a product of two cycles by computing $P^2$ in cycle notation. Since we defined $P = (1,2,3)$ we have $P_2 = P^2 = (1,2,3)(1,2,3)$. By the above definition 
\\
$P_2(1) = (1,2,3)(1,2,3)(1) = (1,2,3)(2) = (3)$, \\
$P_2(2) = (1,2,3)(1,2,3)(2) = (1,2,3)(3) = (1)$, \\
$P_2(3) = (1,2,3)(1,2,3)(3) = (1,2,3)(1) = (2)$. \\
Therefore $P_2 = (1,3,2)$.
If we enumerate the basis states as 
\begin{equation}
1 \equiv |1\rangle \equiv
\begin{bmatrix}
1 \\ 0 \\ 0
\end{bmatrix} ,\;\; 
2 \equiv |2\rangle \equiv
\begin{bmatrix}
0 \\ 1 \\ 0
\end{bmatrix} ,\;\; 
3 \equiv |3\rangle \equiv 
\begin{bmatrix}
0 \\ 0 \\ 1
\end{bmatrix} ,\;\; 
\end{equation}
then with the action described above one can see that $|k\rangle = P_i^{\text{matrix notation}} |j\rangle$ corresponds to
$k=P_i^{\text{cycle notation}} (j)$. \\

In the above notation, the set of $n \times n$ permutation matrices can be seen as groups generated by an $n$-cycle. Any given $n$-cycle \hbox{$\sigma = (a_0,a_1,a_2,a_3,...,a_{n-1}) \in S_n$}, where $S_n$ is the symmetric permutation group, has order $n$. To see why this is so, observe that $\sigma^k(a_0) = a_k$ for $0<k<n$ so its order cannot be less than $n$ and $\sigma^n$ is the identity as $\sigma^n(a_i) = a_i$ for $i \in \{0, n-1\}$. Therefore the group generated by $\sigma$ has $n$ elements. More generally we have: $\sigma^{k}(a_j) = a_{(j+k) \; mod \; n}$.  This implies $\sigma^{k_1}(a_j) \neq \sigma^{k_2}(a_j)$ for $k_1 \neq k_2$. 

Let $P$ be the permutation matrix corresponding to $\sigma$.
Then, $P^{k_1} |a_j\rangle \neq P^{k_2} |a_j\rangle$ for $k_1 \neq k_2$ and basis vector $|a_j \rangle$. Since any row (or column) of a permutation matrix has value 0 at all positions but one, where it has the value 1, this implies that no two permutation matrices generated from $P$ have the same row otherwise that would mean $P^{k_1} |a_j\rangle = P^{k_2} |a_j\rangle$ for some $k_1, k_2, a_j$. 

Next, we will show that for any given matrix entry $(i,j)$ there is at least one permutation matrix $P^k$ having value 1 at $(i,j)$, that is $P^k_{i,j} =1$ for some $k$. Let $|i\rangle$ denote the basis vector which has value 1 at the $i$-th index and 0 at all others. Since these permutation matrices form a group there must be some matrix $P^k$ such that $P^k |j\rangle = |i\rangle$. This however means that $P^k$ has value 1 at $(i,j)$. Combining the two statements above, we find that for any entry $(i,j)$ there is exactly one permutation matrix generated from $P$ that has value 1 at that entry. The diagonal matrices $D_j$ may be used to convert these 1's to any desired value. We have thus shown that by choosing our $P_j$ permutation matrices as $n$-cycles, one can construct arbitrary Hamiltonians $H=\sum_j D_j P_j$. This proof also provides a prescription as to how to explicitly choose the permutations $P_j$. 

\section{Hermiticity of $H$ \label{app:Hermicity_proof}}
Here we show that $H=\sum D_j P_j$ is hermitian if and only if for every index $j$ there is an associated index $j'$ such that $P_j = P_{j'}^{-1}$ and $D_j = D_{j'}^*$ (in general, $j$ and $j'$ may correspond to the same index). 

We first prove the if direction. Let $H$ above be a hermitian matrix. We show that this implies that for every index $j$ there is an associated index $j'$ such that \hbox{$P_j = P_{j'}^{-1}$}. Let $P_j$ be the permutation sending a basis vector $p$ to another basis vector $q$. Thus in `cycle notation' (see Appendix \ref{app:PermutationM}) \hbox{$P_j = \ldots (\ldots p,q \ldots) \ldots$}. Let us assume that $D_j$ is not the zero matrix (otherwise $D_j P_j$ is trivially zero). Case I: Let ${D_i}_{(q,q)} \neq 0$. Since $H$ is hermitian thus there exist $P_{j'} = \ldots (\ldots q,p \ldots) \ldots$ in the decomposition of $H$. But then $P_j P_{j'} =  \ldots (\ldots q,q \ldots) \ldots$ which has to be equal to $\mathbb{1}$ (as otherwise this would imply the existence of a permutation matrix that has a fixed point contradictory to our initial setup). Case II: Let ${D_j}_{(q,q)} = 0$. Since $D_j$ is not identically zero, there exists an element $s$ such that ${D_j}_{(s,s)} \neq 0$. Let us denote by $r$ the element that is sent to $s$ in $P_j$, that is  $P_j= \ldots (\ldots p,q \ldots r,s \ldots) \ldots$. Again since $H$ is hermitian there must exist  $P_{j'} = \ldots (\ldots s,r \ldots) \ldots$ in the decomposition of $H$. But then $P_j P_{j'} =  \ldots (\ldots s,s \ldots) \ldots$ which has to be equal to $\mathbb{1}$. Thus $P_j$ has an inverse $P_{j'}$ in the decomposition of $H$. 

Next, we prove that there can be no two permutation matrices in $H$ with the same nonzero element. In cycle notation, this assertion translates to the assertion that there can be no two distinct permutations that send a basis state to the same basis state. We prove this by contradiction. Let $P_j$ and $P_k$ be two distinct permutations both of which send a basis vector $p$ to $q$. In 'cycle notation' this mean $P_j = \ldots (\ldots p,q \ldots) \ldots$ and $P_k = \ldots (\ldots p,q \ldots) \ldots$ with $P_j \neq P_k$. We showed above that $P_j$ has an inverse $P_{j'}$, that is $P_j P_{j'} = \mathbb{1}$. But $P_{j'} P_k = \ldots (\ldots p,p \ldots) \ldots$ has fixed point and cannot be the identity due to the uniqueness of the inverse. Thus we have reached a contradiction. 

We now prove the full if part. By definition, that $H$ is hermitian implies $\sum D_j P_j = \sum D_j^{*} {P_j}^{-1}$. Let $P_{j'}$ be the inverse of $P_j$ which as we proved should exist in the decomposition with a nonzero weight $D_j$. Equating the right-hand side and the left-hand side of the equality gives $D_j = D_{j'}^*$. 

Proving the other direction is simpler. We assume that in the summation $H=\sum D_j P_j$ there is for every index $j$ an associated index $j'$ such that $P_j = {P_{j'}}^{-1}$ and $D_j = {D_{j'}}^{*}$. Thus $H^{\dagger}=\sum {D_{j'}}^{*} {P_{j'}}^{-1} = \sum {D_j} {P_j} = H$. 
\end{document}